\documentclass[aps,twoside,amsmath,amssymb,longbibliography]{revtex4-1}

\usepackage[paperwidth=21.59cm,paperheight=27.94cm,hmargin=2.54cm,vmargin=2.54cm,asymmetric]{geometry}
\usepackage{graphicx} % include figure files
\usepackage{dcolumn} % align table columns on decimal point
\usepackage{bm} % bold math
\usepackage{hyperref} % include hyperlinks
\usepackage{ragged2e} % justify
\usepackage{setspace} % change line spacing
\usepackage{fancyhdr} % include fancy headers
\usepackage{upgreek} % curly tau
\usepackage{float} % place here only
\usepackage{tikz} % draw blue circle and red cross
\usepackage{gensymb} % degree symbol
\usepackage{mathrsfs} % curly R
\usepackage{color} % text color
\usepackage{stackengine} % hyperlinks in images
\usepackage{xcolor,soul} % highlight
\usepackage{mathtools}
\usepackage{bigints}
\usepackage{relsize}

\hypersetup{colorlinks,urlcolor=blue,linkcolor=blue,citecolor=blue} % blue
\newcommand\Int{\bigintsss}
\fancypagestyle{tstyle}{\chead{}} % title page header
\newcommand{\tc}{\textcolor}

\fancypagestyle{prfstyle}{\fancyhf{}
\fancyhead[CO]{Transient Growth Analysis of Oblique SWBLI at Mach $5.92$}
\fancyhead[CE]{Dwivedi, Hildebrand, Nichols, Candler, and Jovanovi\'c}}
\newlength\myindention

\begin{document}

\title{Transient growth analysis of oblique shock wave/boundary-layer \\[0.25cm]interactions at Mach $5.92$}

\author{Anubhav Dwivedi}
\email[E-mail:]{dwive016@umn.edu}
\affiliation{Department of Aerospace Engineering and Mechanics, University of Minnesota, 110 Union Street SE, Minneapolis, Minnesota 55455-0153, USA}
\author{Nathaniel Hildebrand}
\email[E-mail:]{hilde115@umn.edu}
\affiliation{Department of Aerospace Engineering and Mechanics, University of Minnesota, 110 Union Street SE, Minneapolis, Minnesota 55455-0153, USA}
\author{Joseph W.\ Nichols} 
\email[E-mail:]{jwn@umn.edu}
\affiliation{Department of Aerospace Engineering and Mechanics, University of Minnesota, 110 Union Street SE, Minneapolis, Minnesota 55455-0153, USA}
\author{Graham V.\ Candler}
\email[E-mail:]{candler@umn.edu}
\affiliation{Department of Aerospace Engineering and Mechanics, University of Minnesota, 110 Union Street SE, Minneapolis, Minnesota 55455-0153, USA}
\author{Mihailo R.\ Jovanovi\'c}
\email[E-mail:]{mihailo@usc.edu}
\affiliation{Ming Hsieh Department of Electrical and Computer Engineering, University of Southern California, 3740 McClintock Avenue, Los Angeles, California 90089-2560, USA}

% \vspace*{0.5cm}

\begin{abstract}
\tc{black}{We study physical mechanisms that trigger transient growth in a high-speed spatially-developing laminar boundary layer that interacts with an oblique shock wave. We utilize an approach based on power-iteration, with the global forward and adjoint linearized equations, to quantify the transient growth in compressible boundary layers with flow separation. For a Mach $5.92$ boundary layer with no oblique shock wave, we show that the dominant transient response consists of oblique waves, which arise from the inviscid Orr mechanism, the lift-up effect, and the first-mode instability. We also demonstrate that the presence of the oblique shock wave significantly increases transient growth over short time intervals through a mechanism that is not related to a slowly growing global instability. The resulting response takes the form of spanwise periodic streamwise elongated streaks and our analysis of the linearized inviscid transport equations shows that base flow deceleration near the reattachment location contributes to their amplification. The large transient growth of streamwise streaks demonstrates the importance of non-modal effects in the amplification of flow perturbations and identifies a route for the emergence of similar spatial structures in transitional hypersonic flows with shock wave/boundary-layers interaction.}
\end{abstract}

\setstretch{1.}

\maketitle\thispagestyle{tstyle}

\section{INTRODUCTION}
\vspace*{-2ex}

\tc{black}{Shock wave/boundary-layer interactions (SWBLI) are commonly encountered in high-speed flows over complex geometries that involve intakes, control surfaces, and junctions. An oblique shock wave impinging on the flat-plate boundary layer is a canonical setup used to investigate the resulting flow separation and reattachment in SWBLI~\cite{Currao2019,Sandham}. Despite the spanwise homogeneity of the flat-plate geometry, experiments~\cite{Currao2019} and numerical simulations~\cite{Sandham,Pagella1} demonstrate the emergence of three-dimensional streamwise streaky flow structures near the reattachment location before transition to turbulence.}

\tc{black}{Recent studies of two-dimensional laminar SWBLI~\cite{Chuvakhov2020,CaoKliHer2019} show that such configurations can significantly amplify external three-dimensional disturbances. Previous analyses focused on amplification of flow perturbations that arise from streamwise curvature of streamlines in the region where the separated flow reattaches to the wall~\cite{navarro2005,priebe2016low}. Recent experimental observations also highlight the presence of three-dimensional structures inside the recirculation bubble~\cite{Zhuang2017}. In this paper, we consider the role of the entire recirculation bubble in the amplification of flow perturbations.}

\tc{black}{Several studies investigated the effect of the separated flow by carrying out global stability analysis~\cite{Theofilis, TheofilisARFM}. These references show that 2D SWBLI that arise from impinging oblique shock waves~\cite{Hildebrand, Robinet} and compression corners~\cite{Sidharth} exhibit an intrinsic 3D global instability inside the separation bubble as the strength of  interactions increase. However, in contrast to recent experimental observations~\cite{Chuvakhov2017}, the wavelength predicted by global stability analysis scales with the length of the recirculation zone~\cite{gs2017global}, thereby suggesting that other amplification mechanisms may play a more prominent role.}

\tc{black}{In the absence of global instabilities, previous studies~\cite{Butler, Andersson, Jovanovic} showed that non-modal amplification can lead to subcritical transition. Despite the asymptotic decay of the flow perturbations, non-orthogonality of the associated eigenmodes can induce significant transient amplification and trigger non-linear interactions that ultimately break down into turbulence~\cite{Schmid1}. This amplification is closely related to high sensitivity of laminar flow to external disturbances~\cite{mj-phd04,Jovanovic,Schmid2,jovARFM20} and it can be quantified using frequency domain input-output analysis. Recent work on SWBLI over compression ramps~\cite{Dwivedi3} and double wedges~\cite{Dwivedi2} demonstrated significant amplification of steady upstream vortical disturbances around globally stable laminar flow.}

\tc{black}{Another common approach to quantifying non-modal amplification is through computation of maximum growth of perturbations over a given interval in time or space~\cite{Schmid1}. Most previous studies of compressible flows evaluate the initial conditions, which result in optimal temporal growth at a given spatial location~\cite{Hanifi, Bitter1, Bitter2}. For perturbations of a given frequency, a similar approach based on the linearized boundary layer equations provides a method for quantifying the optimal spatial amplification in spatially evolving flows~\cite{Luchini2, Tumin2, Zuccher, Tempelmann2, Paredes1}. For SWBLI, this methodology was recently utilized to evaluate the spatial growth of streamwise streaks in the reattaching boundary layers in the absence of flow recirculation~\cite{Dwivedi1} .}

\tc{black}{In the present work, we utilize the global linear system to evaluate the optimal spatio-temporal growth of canonical SWBLI configuration in which an oblique shock wave impinges on a flat plate boundary layer at Mach $5.92$. We first demonstrate the validity of our approach by evaluating amplification mechanisms in a spatially-evolving hypersonic boundary layer. In spite of the presence of a slowly growing global instability in the oblique SWBLI, we show that significant transient growth arises over short time intervals through a mechanism that is not related to global instability. The dominant response takes the form of streamwise streaks, which are ubiquitous in experiments and numerical simulations.  To uncover physical mechanisms responsible for this significant increase in transient growth, we analyze the inviscid transport of the optimal wavepacket in the flow. Similar to SWBLI setups without a global instability~\cite{Dwivedi3}, our analysis demonstrates that the base flow deceleration plays a dominant role in amplification of streamwise-streaky flow structures.}

\tc{black}{Our presentation is organized as follows. In Section~\ref{sec.problem}, we formulate the problem, describe the governing equations and the numerical method, and summarize the approach that we use to compute the spatial structure of the optimal initial conditions and the resulting temporal growth envelopes. In Section~\ref{nonparallel}, we investigate the importance of convective instabilities in a Mach $5.92$ spatially-developing boundary layer and, in Section~\ref{sbli}, we examine how the formation of a recirculation bubble in SWBLI changes the spatio-temporal response of the linearized flow equations. In Section \ref{mech}, we conduct a wavepacket analysis to show the emergence of oblique waves in the absence of SWBLI and streamwise streaks in the presence of the shock. We also uncover physical mechanisms responsible for transient growth by examining the inviscid transport equation. We conclude our presentation in Section~\ref{conclude}.}

	\vspace*{-4ex}
\section{PROBLEM FORMULATION}
	\label{sec.problem}

	\vspace*{-2ex}
\subsection{Flow configuration}

\begin{figure}%[htb] % order of placement preference: here, top, bottom
	% \setstretch{1.}
	\includegraphics[width=16cm]{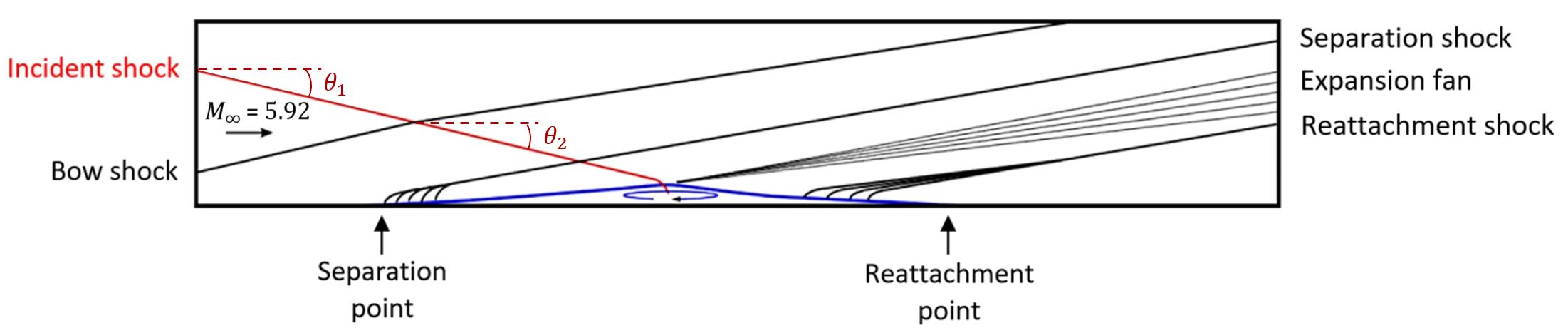}%\vspace{0.2cm}
	\caption{Schematic of an oblique shock wave (red) impinging on a Mach 5.92 boundary layer, adapted from~\cite{Hildebrand}. The adverse pressure gradient associated with the incident shock causes the boundary layer to separate from the wall, forming a recirculation bubble (blue). Here, $\theta_1$ is the initial angle of the incident shock, while $\theta_2$ denotes the shock angle after the incident shock interacts with the bow shock.}
	\label{domain}
	\vspace*{-0.4cm}
\end{figure}

\vspace*{-2ex}
A spatially-developing boundary layer and an oblique shock-wave/boundary-layer interaction are considered in this paper. For both of these flow configurations, the free-stream conditions match experiments performed in the ACE Hypersonic Wind Tunnel at Texas A\&M University \cite{Semper}. As shown in Figure~\ref{domain}, a laminar boundary layer enters at the left boundary and travels over a flat plate situated along the bottom boundary. \tc{black}{We examine} a unit Reynolds number $4.6\times10^{6}$ m$^{-1}$ \tc{black}{or} $Re=\rho_{\infty}u_\infty\delta^*_\text{in}/\mu_\infty=9660$ based on the undisturbed boundary-layer displacement thickness $\delta^*_\text{in}=2.1$ mm at the left boundary. Here, $\rho_\infty$, $u_\infty$, and $\mu_\infty$ denote the free-stream density, velocity, and dynamic viscosity, respectively. A bow shock that persists throughout the domain is produced by the leading edge of the flat plate (which is upstream of the left boundary). For the SWBLI, which is depicted in Figure~\ref{domain}, an incident oblique shock wave enters the computational domain through the left boundary well above the bow shock and the boundary layer. \tc{black}{This incident shock propagates at an angle $\theta_1$ and its interaction with the bow shock changes the angle to $\theta_2$. Upstream of the bow and incident shocks, the free-stream Mach number, temperature, and pressure are $M_\infty=5.92$, $T_\infty=53.06$ K, and $p_\infty=308.2$ Pa. After the shock/shock interaction, the incident oblique shock wave strikes the boundary layer, thereby causing separation from the wall and formation of a recirculation bubble.}

	\vspace*{-4ex}
\subsection{Governing equations}

	\vspace*{-2ex}
We use the compressible Navier-Stokes equations to model the dynamics of a spatially-developing boundary layer and an oblique SWBLI. These equations govern the spatio-temporal evolution of the state $[ \, p ~\, \bm{u}^\mathrm{T}  ~\, s \,]^\mathrm{T}$, where $p$, $\bm{u}$, and $s$ are the nondimensional pressure, velocity, and entropy, respectively~\cite{Sesterhenn}. After nondimensionalization with respect to the displacement thickness $\delta^*_\text{in}$, free-stream velocity $u_\infty$, density $\rho_\infty$, and temperature $T_\infty$, the equations take the form
\begin{equation}
\begin{array}{rcl}
	\dfrac{\partial{p}}{\partial{t}} \; + \; \bm{u\cdot\nabla}p \; + \; \rho{a^2}\bm{\nabla\cdot{u}}
	& = & 
	\dfrac{1}{Re} \left( \dfrac{1}{M_\infty^2Pr} \, \bm{\nabla\cdot}(\mu\bm{\nabla}T) \; + \; (\gamma-1)\phi \right),
	\\[0.25cm]
	\dfrac{\partial\bm{u}}{\partial{t}} \; + \; \dfrac{1}{\rho} \, \bm{\nabla}p+\bm{u\cdot\nabla{u}}
	& = & 
	\dfrac{1}{Re} \, \dfrac{1}{\rho} \, \bm{\nabla\cdot\uptau},
	\\[0.25cm]
	\dfrac{\partial{s}}{\partial{t}} \; + \; \bm{u\cdot\nabla}s
	& = &
	\dfrac{1}{Re}\dfrac{1}{\rho{T}}
	\left(
	\dfrac{1}{(\gamma-1)M_\infty^2Pr} \, \bm{\nabla\cdot}(\mu\bm{\nabla}T) \; + \; \phi
	\right).
\end{array}
\label{NS}
\end{equation}
We define the free-stream Mach number as $M_\infty=U_\infty/a_\infty$, where $a_\infty=\sqrt{\gamma{p_\infty/}\rho_\infty}$ denotes the free-stream speed of sound. The equation of state for an ideal fluid is $\gamma{M_\infty^2}p =\rho{T}$. Furthermore, $\gamma=1.4$ is the assumed constant ratio of specific heats. We define entropy as $s=\ln(T)/[(\gamma-1)M_\infty^2]-\ln(p)/(\gamma{M_\infty^2})$ so that $s=0$ when $p=1$ and $T=1$~\cite{Nichols2}. 

The viscous stress tensor $\bm{\uptau}$ is written in terms of the identity matrix $\bm{I}$, velocity vector $\bm{u}$, and dynamic viscosity $\mu$ to yield 
\begin{equation}
	\bm{\uptau} 
	\; = \; 
	\mu
	\left( \bm{\nabla{u}} \; + \; (\bm{\nabla{u}})^\mathrm{T} \; - \; \dfrac{2}{3}(\bm{\nabla\cdot{u}})\bm{I} \right).
\end{equation}
We define the viscous dissipation as $\phi= (1/\mu) \, \bm{\uptau \! : \! \nabla{u}}$, where the operator $\bm{:}$ represents a scalar product of two tensors, $\bm{A:B}=\mathrm{trace} \, (\bm{A}^\mathrm{T}\bm{B})$. Further, the second viscosity coefficient is set to $\lambda=-2\mu/3$. In order to compute the dynamic viscosity $\mu$, Sutherland's law is used with $T_s=110.3~\mathrm{K}$,
\begin{equation}
	\mu(T) 
	\; = \; 
	T^{3/2} \, \dfrac{1 \, + \, T_s/T_\infty}{T \, + \, T_s/T_\infty}.
\end{equation}
The Prandtl number is set to a constant value $Pr=c_p\mu/\kappa=0.72$, where $\kappa$ and $c_p$ are the coefficients of heat conductivity and specific heat at constant pressure, respectively. 

To study the behavior of small fluctuations around various base flows, System~\eqref{NS} is linearized by decomposing the state into a sum of the steady $\bar{\bm{q}}$ and fluctuating $\bm{q}$ parts. The linearized Navier-Stokes (LNS) equations take the following form~\cite{Hildebrand_CITE},
\begin{equation}
	\dfrac{\partial\bm{q}}{\partial{t}} 
	\; = \; 
	A \, \bm{q}
	~~
	\Rightarrow
	~~
	{\bm{q}(t)} 
	\; = \; 
	e^{At} \, \bm{q}(0),
\label{LNS}
\end{equation}
where $A$ is the Jacobian operator resulting from the linearization of System~\eqref{NS} around the base state $\bar{\bm{q}}$. As discussed below, the transient growth analysis also requires the adjoint of the Jacobian operator, $A^\text{H}$, \tc{black}{where we compute $A^\text{H}$ using a discrete approach. For a detailed discussion of adjoint operators see~\cite{Luchini3}.}

	\vspace*{-4ex}
\subsection{Power iteration}
\label{adjloopsec}

	\vspace*{-2ex}
We employ an iterative approach, based on power iteration~\cite{Schmid2}, to conduct the transient growth analysis. For compressible flows, the fluctuations' energy is defined as~\cite{Chu,Hanifi},
\begin{equation}
	E 
	\; = \; 
	\dfrac{\bar\rho{u}_i'u_i'^*}{2}
	\; + \; 
	\dfrac{M_\infty^2|p'|^2}{2}
	\; + \;
	\dfrac{(\gamma-1)M_\infty^2|s'|^2}{2},
	\label{energy}
\end{equation}
where $^*$ denotes the complex conjugate.  This expression is obtained by eliminating conservative compression work transfer terms and \tc{black}{the energy is induced by the inner product $(\bm{q}_1,\bm{q}_2)_E=\int \bm{q}_2^\mathrm{H}W\bm{q}_1\,\mathrm{d}x \, \mathrm{d}y$, where the weighting matrix is given by $W = (1/2) \, \mathrm{diag}[M_\infty^2,\bar\rho,\bar\rho,\bar\rho,(\gamma-1)M_\infty^2]$. The energy growth at time $t$ is determined~by,} 
\begin{equation}
	G(t) 
	\; = \;
	\max\limits_{\bm{q}(0) \, \neq \, 0}
	\;
	\dfrac{||\bm{q}(t)||_E^2}{||\bm{q}(0)||_E^2},
	\label{gain}
\end{equation}
\tc{black}{where $\bm{q}(0)$ and $\bm{q}(t)$ represent the initial and final states. Since $\bm{q}(0)$ depends on the time interval on which optimization is performed, $G(t)$ determines an envelope of all possible optimal responses.
For a given interval $(0, t]$, we compute the maximum transient growth by alternating between integration of the governing equations forward in time and integration of the adjoint equations backward in time~[\onlinecite{Schmid2},\onlinecite{Guegan}]. Starting from a random initial condition, this approach converges to the principal eigenfunction of the direct-adjoint system, which is equivalent to the principal singular function of the direct system in Eq.~\ref{LNS}.} 

	\vspace{-4ex}
\subsection{Numerical method}

	\vspace*{-2ex}
\tc{black}{The base flow computations for the spatially developing boundary layer and the oblique SWBLI are carried out using US3D~\cite{candler2015development}, a compressible finite volume flow solver. We apply symmetry boundary conditions along the side of the domain, thereby ensuring that the base flow remains two dimensional.
For the transient growth calculations, we use the formulation by Sesterhenn~\cite{Sesterhenn} but our numerical method differs from~\cite{Sesterhenn}. We use the fourth-order centered finite differences applied on a stretched mesh to discretize the LNS equations. This yields a large sparse matrix~\cite{Nichols2} and time integration is performed using an implicit first-order Euler method.We verify the resulting temporal evolution using an explicit fourth-order Runge-Kutta method and find that there is almost no difference.} The inversion step is computed via the lower-upper (LU) decomposition of the shifted sparse matrix using the parallel SuperLU package~\cite{Li}. A numerical filter is used to add minor amounts of scale-selective artificial dissipation to damp spurious modes associated with the smallest wavelengths allowed by the mesh \cite{Durran}. We introduce the numerical filter by adding terms of the form $\epsilon(\partial^4\bm{q}/\partial{x^4})$ and $\epsilon(\partial^4\bm{q}/\partial{y^4})$. \tc{black}{The value of $\epsilon$ is selected to be as small as possible (e.g., $\epsilon=0.0125$) in order to minimize the impact of this artificial dissipation while maintaining numerical stability.} Furthermore, we check that our results are not sensitive to this choice by repeating our calculations with different values of $\epsilon$. Sponge layers are employed at the top, left, and right boundaries of the nonparallel boundary layer and oblique SWBLI to absorb outgoing information with minimal reflection \cite{Mani}.

	\vspace*{-4ex}
\section{TRANSIENT GROWTH ANALYSIS}
	\label{sec.tg}

	\vspace*{-2ex}
\subsection{Spatially-developing boundary layer}
\label{nonparallel} 

	\vspace*{-2ex}
\tc{black}{In our previous study~\cite{Hildebrand_CITE}, we verified our numerical method for computing the optimal transient growth of parallel high-speed compressible boundary layers~\cite{Bitter1,Hanifi}. The largest transient response in such flows is triggered by the lift-up mechanism~\cite{Landahl1,Landahl2} similar to that observed in the low-speed incompressible wall-bounded shear flows~\cite{Schmid1,jovARFM20}. Acoustic disturbances are also significantly amplified in spatially developing boundary layers~\cite{Bitter2} and we utilize global transient growth analysis to examine the importance of different amplification mechanisms.}

\begin{figure}%[htb] % order of placement preference: here, top, bottom
	% \setstretch{1.}
	\includegraphics[width=15.3cm]{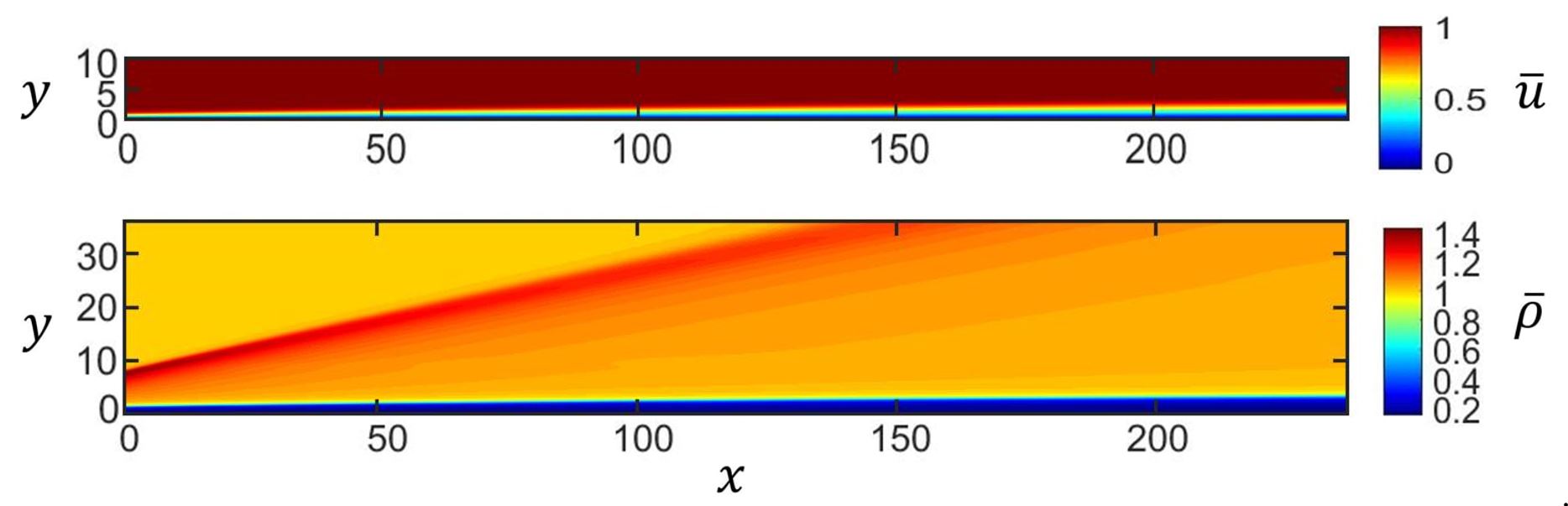}
	\caption{Base flow contours of nondimensional streamwise velocity and density for a Mach 5.92 spatially-developing boundary layer with $Re_{\delta^*_\text{in}}=9660$, adapted from~\cite{Hildebrand_CITE}.}
	\label{base_xsdbl}
		\vspace*{0.1cm}
\end{figure}

\tc{black}{We use US3D to compute the steady two-dimensional flow over an adiabatic wall. The free-stream conditions presented in the previous section correspond to that of the SWBLI without the shock. The numerical domain extends 235$\delta^*_\text{in}$ in the streamwise direction and 36$\delta^*_\text{in}$ in the wall-normal direction.} We utilize a Cartesian mesh to discretize this domain. The mesh is stretched in the wall-normal direction with $y^+=0.6$, and it is uniformly spaced in the streamwise direction. A total of $n_x=500$ and $n_y=420$ grid points are used to resolve this domain in the streamwise and wall-normal directions, respectively. \tc{black}{At the inflow, we impose a boundary-layer profile computed in an earlier study~\cite{Shrestha}. Furthermore, we impose a characteristic-based supersonic outlet boundary condition along the top and right edges of the domain. This boundary condition allows the shock waves to exit the domain without reflection.} The base flow simulations are carried until the numerical residual is of the order of machine precision. \tc{black}{Figure~\ref{base_xsdbl} shows base flow contours of the spatially-developing boundary layer in the absence of the incident oblique shock.}

\begin{figure}[htb] % order of placement preference: here, top, bottom
	% \setstretch{1.}
	\includegraphics[width=14.5cm]{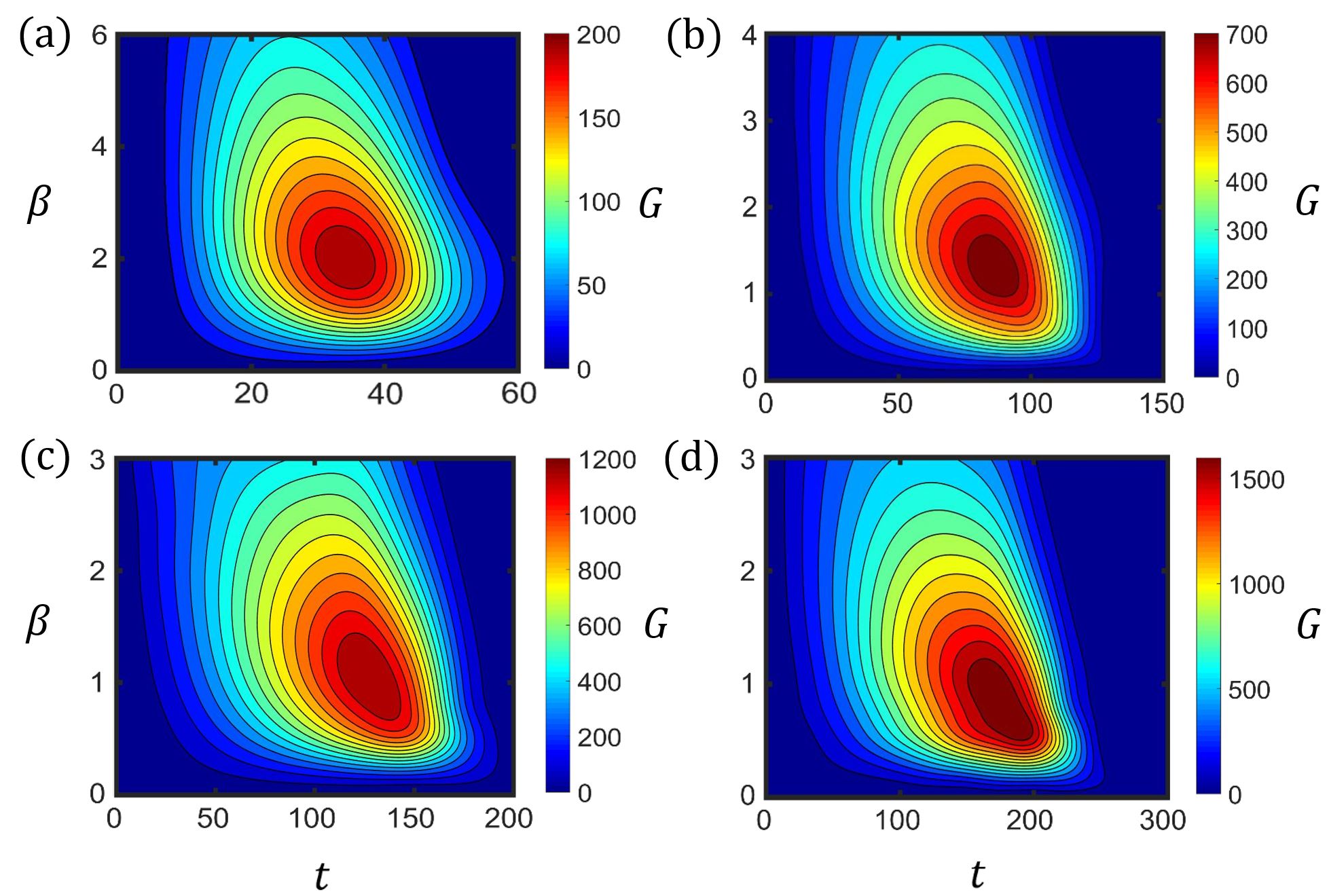}%\vspace{0.2cm}
	\caption{Contour plots of the transient growth $G(\beta,t)$ of a Mach 5.92 spatially-developing boundary layer for streamwise lengths of the domain equal to (a) $59$, (b) $118$, (c) $176$, and (d) $235$. The streamwise lengths correspond to the factors 1/4, 1/2, 3/4, and 1 of the original length. These plots illustrate convective instability of the spatially-developing boundary layer.}
	\label{grow_xsdbl}
\end{figure}

\begin{figure}%[htb] % order of placement preference: here, top, bottom
	% \setstretch{1.}
	\includegraphics[width=13.cm]{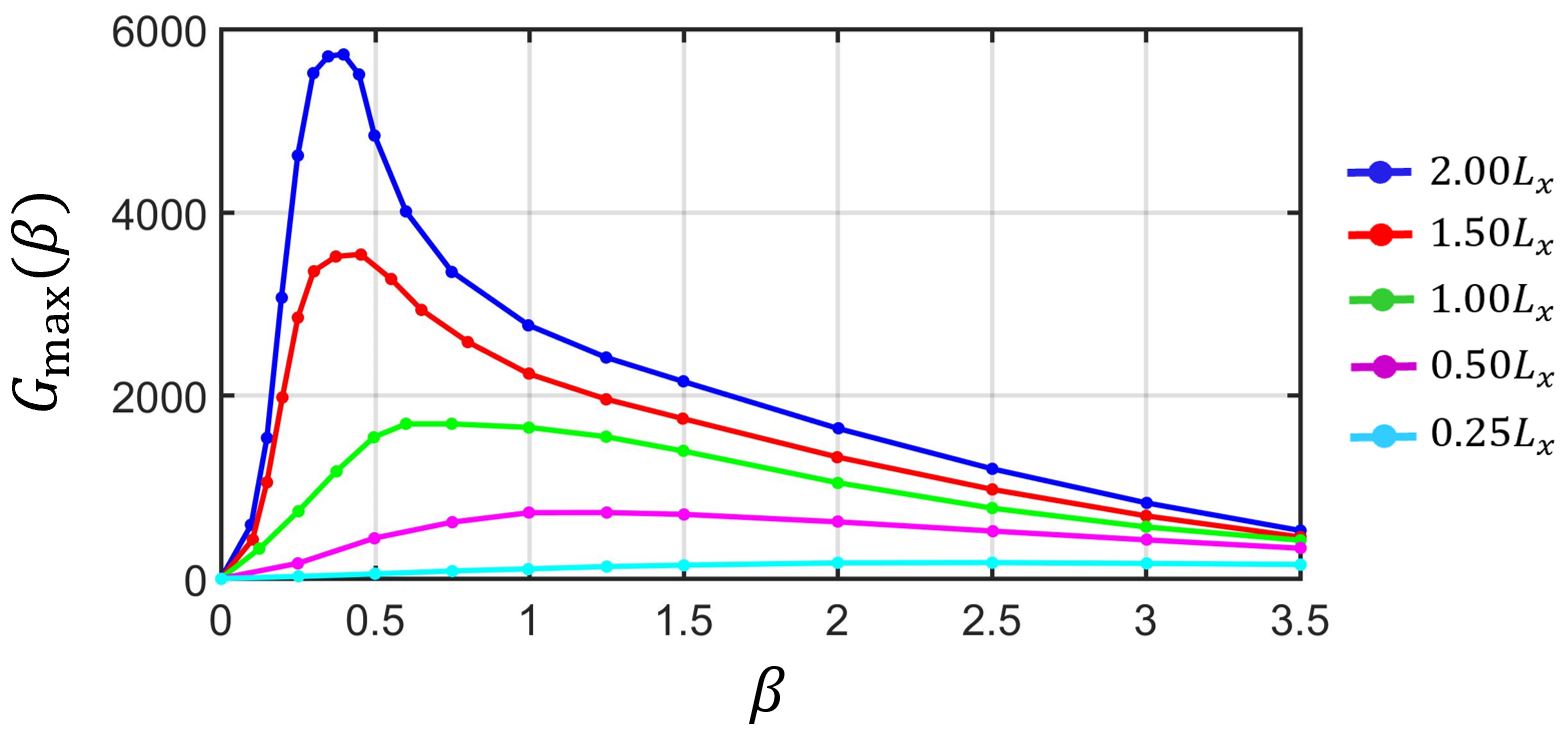}
	\caption{Maximum transient energy growth versus nondimensional spanwise wavenumber of a Mach 5.92 spatially-developing boundary layer with five different streamwise extents for $L_x=235$.}
	\label{beta_xsdbl}
			\vspace*{-0.4cm}
\end{figure}

\tc{black}{The optimal transient growth of the spatially-developing boundary layer is computed by setting all perturbations at the top, left, and right boundaries to zero with the help of three sponge layers~\cite{Mani}. The iterative process is initialized using a random flow field with unit energy. Figure~\ref{grow_xsdbl} shows the transient growth as a function of $t$, nondimensional spanwise wavenumber $\beta$, and length $L_x$ of the streamwise domain. The transient growth becomes larger as the streamwise extent of the domain increases, thereby indicating that the spatially-developing boundary layer is convectively unstable. In other words, even though the linearized system is globally stable, localized upstream perturbations grow substantially as they convect downstream with the flow~\cite{Cossu,Nichols4}. Figure~\ref{beta_xsdbl} displays the dependence on the spanwise wavenumber of the largest transient energy growth, $G_{\max} = \max_t G (t)$, for five different streamwise domain lengths. The largest transient growth occurs for the longest domain and is equal to $5.63\times10^3$ for $\beta=0.38$, $t=350$, and $L_x = 470$. }

\begin{figure}[htb] % order of placement preference: here, top, bottom
	% \setstretch{1.}
	\includegraphics[width=11.5cm]{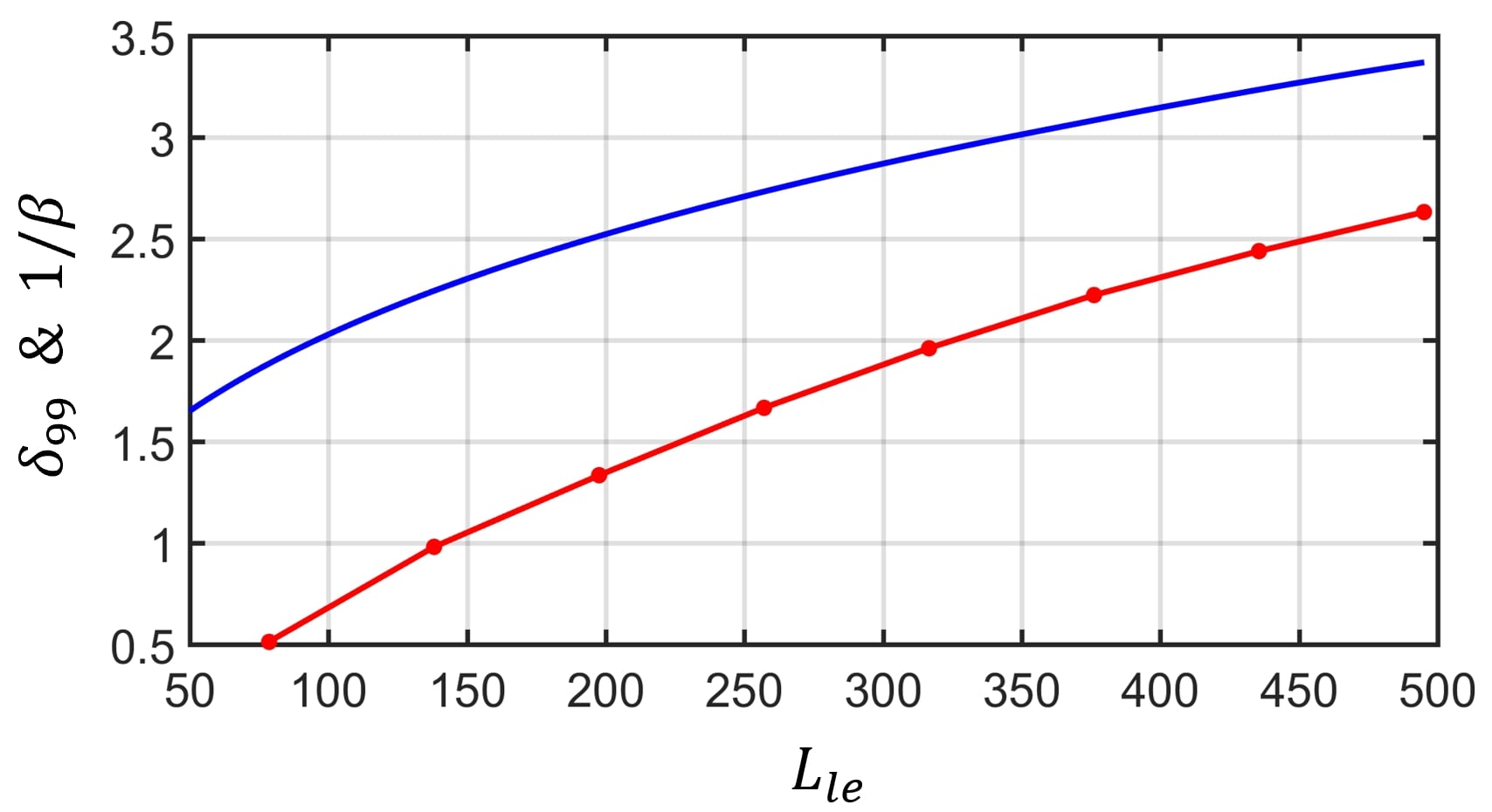}
	\caption{Boundary-layer thickness (solid blue) and the inverse spanwise wavenumber (solid/dots red) corresponding to the maximum transient growth versus the streamwise length of the domain from the leading edge of the flat plate $L_{le}$. This corresponds to a Mach $5.92$ spatially-developing boundary layer with data from Figures \ref{grow_xsdbl} and \ref{beta_xsdbl}.}
	\label{delta_ibeta}
		\vspace*{-0.25cm}
\end{figure}

\tc{black}{In contrast to the impact of the streamwise domain length, the spanwise wavenumber corresponding to the maximum growth decreases. In fact, figure \ref{delta_ibeta} shows that the spanwise wavelength (which is proportional to the inverse of the spanwise wavenumber) scales with the boundary-layer thickness at the right outlet of the domain. Thus, Figures~\ref{beta_xsdbl} and \ref{delta_ibeta} imply that a significant portion of the transient growth occurs near the streamwise end of the domain, and that this region selects the spanwise wavelength that supports the largest growth. This is a consequence of the square-root growth of the boundary layer in the presence of convective instabilities: the spanwise length scale of the local instability near the end of the domain dictates the global response because of the slow spatial development of the boundary layer~\cite{Bitter2}.}

\begin{figure}%[htb] % order of placement preference: here, top, bottom
	% \setstretch{1.}
	\centering\includegraphics[width=14.5cm]{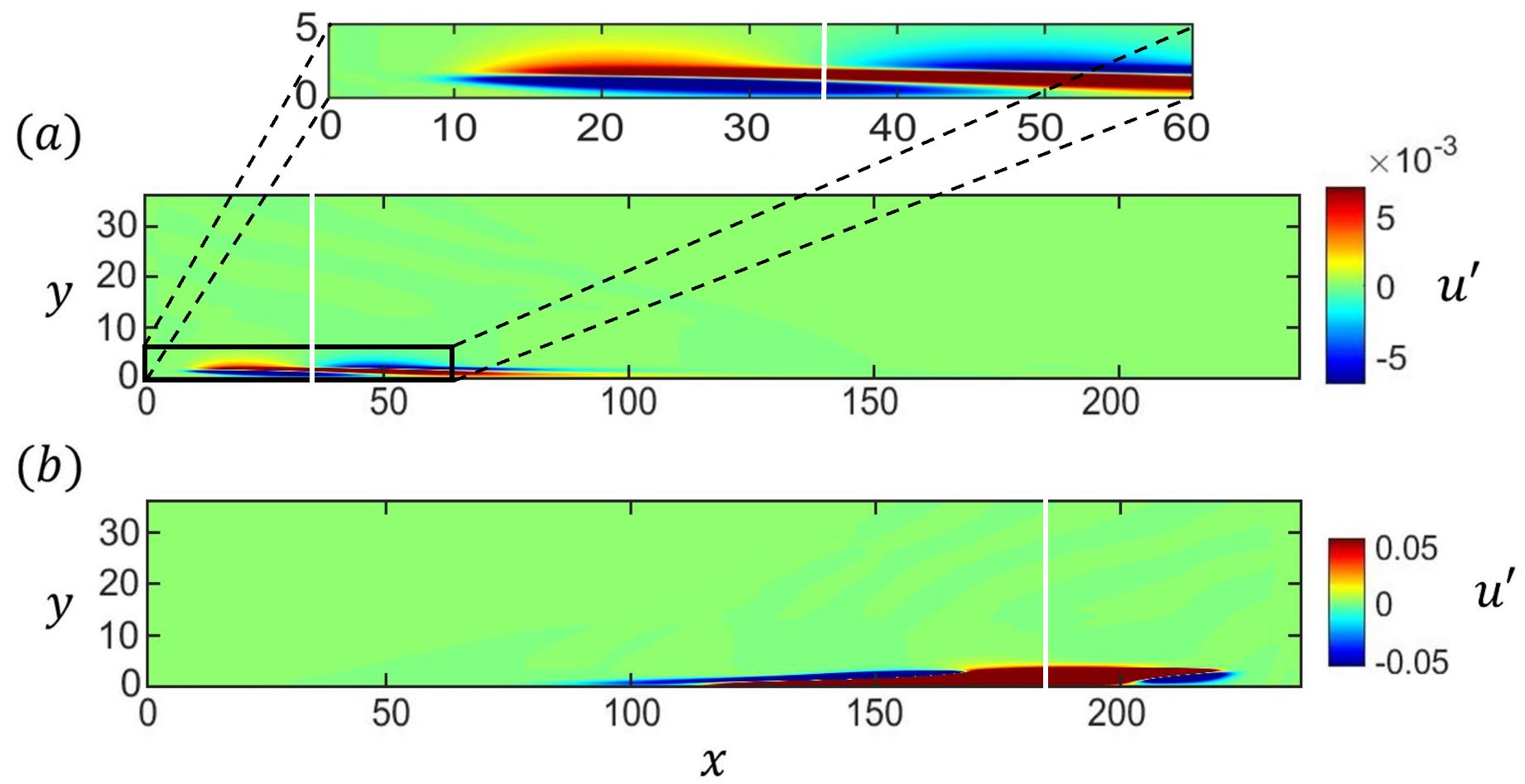}
		\caption{The real part of the normalized streamwise velocity component of the optimal (a) initial and (b) final states of a Mach 5.92 spatially-developing boundary layer with $\beta=0.6$. These states correspond to the peak in Figure \ref{beta_xsdbl} for a streamwise extent of $L_{x} = 235$. The white lines indicate streamwise locations where the wall-normal profiles in Figure \ref{lift_xsdbl} are extracted.}
	\label{io_xsdbl}
			\vspace*{-0.6cm}
\end{figure}

\tc{black}{In Figure \ref{io_xsdbl}, we illustrate the real part of the normalized streamwise velocity component of the optimal initial and final states of the spatially-developing boundary layer with $L_x=235$ and $\beta=0.6$.} The initial state is comprised of the streamwise elongated structures near the left inlet that are tilted against the mean shear of the boundary layer. We note that these streamwise \tc{black}{structures} have a shallow angle because of the high-speed nature of the flow. \tc{black}{As time increases, these tilted streaks start to align with the mean shear thereby causing substantial growth via the inviscid Orr mechanism~\cite{Schmid1}. In wall-bounded shear flows, these flow structures grow rapidly and robustly~\cite{Butler}.} The final state in Figure \ref{io_xsdbl}(b) also \tc{black}{consists of tilted streamwise perturbations along the mean shear} that extend to the right outlet.

\begin{figure}%[htb] % order of placement preference: here, top, bottom
	% \setstretch{1.}
	\centering\includegraphics[width=16cm]{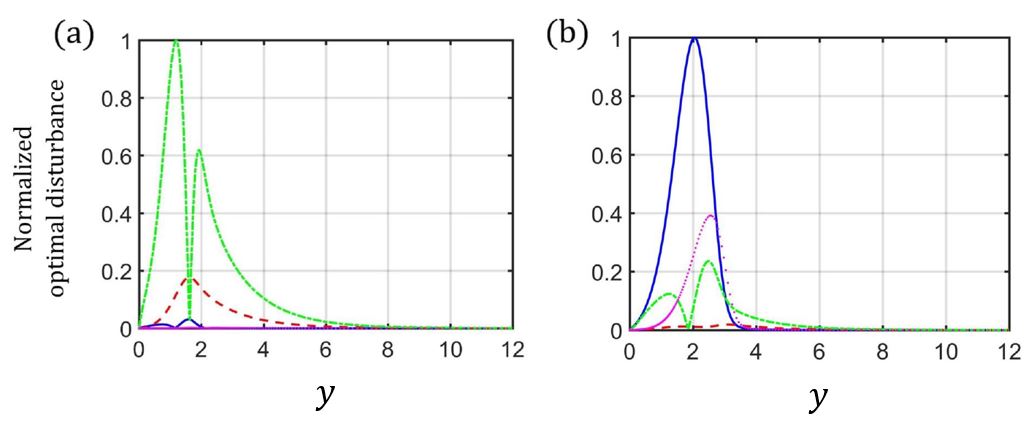}
	\caption{Normalized wall-normal profiles of the optimal (a) initial and (b) final states of a Mach 5.92 spatially-developing boundary layer with $\beta=0.6$. The absolute value of the streamwise (solid blue), wall-normal (dashed red), and spanwise (dash-dot green) velocity components are plotted for each state along with the entropy (dots magenta). The initial and final states are extracted at $x=35$ and $x=185$, respectively.}
			\vspace*{-0.25cm}
	\label{lift_xsdbl}
\end{figure}

To examine whether the lift-up effect~\cite{Landahl1} contributes to the transient response of the spatially-developing boundary layer, we extract wall-normal profiles of the entropy along with the streamwise, wall-normal, and spanwise velocity components of the optimal initial and final perturbations at $x=35$ and $x=185$, respectively. \tc{black}{In Figure \ref{lift_xsdbl}, the perturbations are normalized to have a unit maxima. We see that the wall-normal and spanwise velocities significantly contribute to the optimal initial condition, while the contributions of the streamwise velocity and the entropy are negligible. At the final time, however, the streamwise velocity is much larger than the other velocity components. This suggests that the lift-up mechanism indeed contributes to the optimal transient growth of the spatially-developing boundary layer. While initial streamwise vortices decay in time, streamwise elongated flow structures experience rapid and robust growth~\cite{Bitter1}. Along with the streamwise velocity perturbations, in Figure~\ref{lift_xsdbl} we also notice that the entropy grows substantially with time. This is in concert with the observation that the lift-up effect has a strong  impact on the optimal transient growth~\cite{Tumin2}. In summary, we conclude that it is most effective to excite the wall-normal and spanwise velocity components and that the most energy is carried by the streamwise velocity and the entropy perturbations.}

\begin{figure}%[htb] % order of placement preference: here, top, bottom
	% \setstretch{1.}
	\centering\includegraphics[width=15cm]{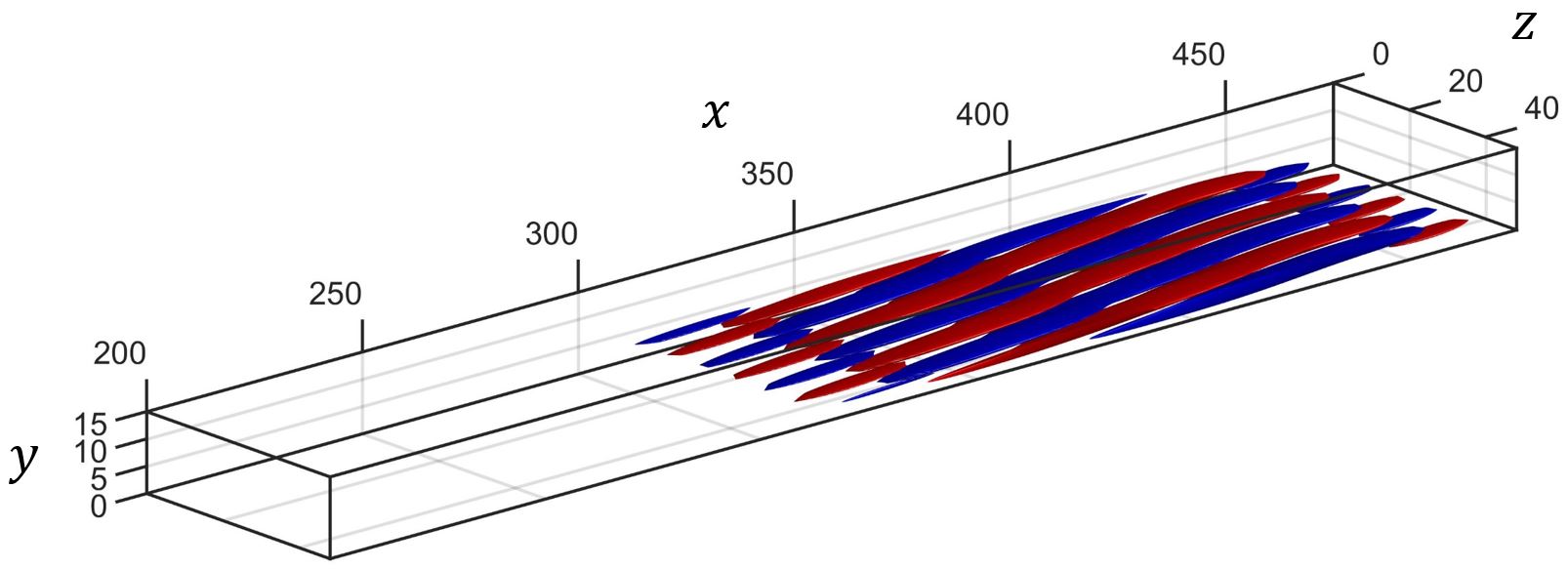}
	\caption{Optimal final state in 3D of a Mach 5.92 spatially-developing boundary layer with $\beta=0.38$. The streamwise domain length is set to $2L_x=470$. Isosurface contours represent the normalized streamwise velocity perturbation, where the red and blue contours indicate positive and negative velocities, respectively.}
	\label{oblique_streaks}
		\vspace*{-0.5cm}
\end{figure}

	\tc{black}{
Similar to the incompressible three-dimensional boundary layers~\cite{Tempelmann1}, this section demonstrates the importance of both the inviscid Orr and the lift-up mechanisms in spatially-developing compressible boundary layers. However, as shown in Figures \ref{grow_xsdbl} and \ref{beta_xsdbl}, the transient growth becomes larger as the streamwise extent of the domain increases. This observation suggests the presence of a convective instability. In Figure \ref{oblique_streaks}, we show the optimal three-dimensional final state of the Mach $5.92$ spatially-developing boundary layer in a longer domain with $\beta=0.38$ and $2L_x=470$. The final state consists of oblique streamwise structures that are reminiscent of first-mode instability. To corroborate the presence of the first-mode instability in the optimal transient response of the spatially-developing boundary layer, we closely examine a wavepacket response in Section~\ref{mech} (see Figure \ref{group_velocity}). Thus, in addition to the contributions arising from the inviscid Orr and lift-up effects the optimal transient response of the spatially-developing boundary layer also arises from the first-mode instability.
	}

	\vspace{-4ex}
\subsection{Shock-wave/boundary-layer interaction}
\label{sbli}

	\vspace{-2ex}
As shown in the previous section, a flat-plate boundary layer at Mach 5.92 supports significant transient growth. In this section, we examine the impact of an oblique SWBLI, at the same conditions. We utilize the US3D compressible flow solver \cite{Candler} to compute a steady-state two-dimensional base flow of an oblique shock-wave/laminar-boundary-layer interaction at Mach $5.92$. The incident oblique shock wave is introduced by modifying the inlet boundary-layer profile so that the Rankine-Hugoniot jump conditions are satisfied at the point where it enters the domain. We select this point so that the incident shock impinges upon the wall at a distance of $119\delta^*_\text{in}$ from the leading edge. \tc{black}{In this study, we are only interested in an oblique shock angle of $\theta_1=13\degree$, which changes to an angle of $\theta_2=12.89\degree$ after relatively weak interaction with the upstream bow shock. All the other boundary conditions are identical to those described in Section~\ref{nonparallel}.} 

\begin{figure}%[htb] % order of placement preference: here, top, bottom
	% \setstretch{1.}
	\includegraphics[width=0.8\textwidth]{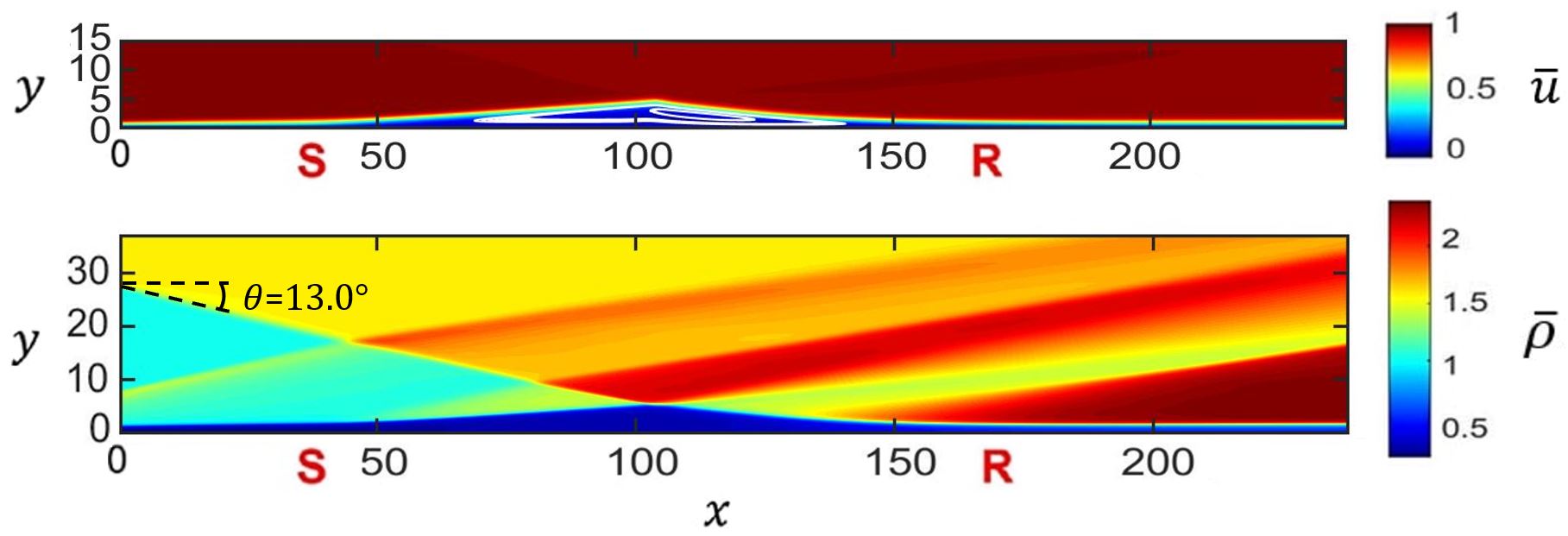}
	\caption{Contour plots of nondimensional streamwise velocity and density for a Mach 5.92 SWBLI with an incident shock angle of $\theta_1=13\degree$ at $Re_{\delta^*_\text{in}}=9660$, adapted from~\cite{Hildebrand_CITE}. Here, \color{red}\textbf{S}\color{black}\ and \color{red}\textbf{R}\color{black}\ denote the separation and reattachment points, respectively. The white contour lines indicate streamlines inside the recirculation bubble.}
	\label{base_xsbli}
			\vspace*{-0.6cm}
\end{figure}

The domain extends $235\delta^*_\text{in}$ and $36\delta^*_\text{in}$ in the streamwise and wall-normal directions, respectively. This domain is discretized by a Cartesian mesh that is nonuniformly spaced in the wall-normal direction with $y^+=0.6$ and uniformly spaced in the streamwise direction. A total of $n_x=998$ and $n_y=450$ grid points are used to resolve this domain. Figure \ref{base_xsbli} illustrates color plots of nondimensional streamwise velocity and density perturbations around the SWBLI base flow. \tc{black}{All length scales are non-dimensionalized with the boundary layer thickness at the inflow $\delta^*_\text{in}$. The incident oblique shock wave causes the boundary layer to separate from the wall at $x\approx50$, leading to the formation of a separation shock and recirculation bubble.} At the recirculation bubble apex, an expansion fan forms and extends up into the free-stream. Moreover, at $x\approx155$, the flow reattaches to the wall and compression waves coalesce to form a second \mbox{reflected shock wave.} 

\begin{figure}%[htb] % order of placement preference: here, top, bottom
	% \setstretch{1.}
	\centering\includegraphics[width=0.78\textwidth]{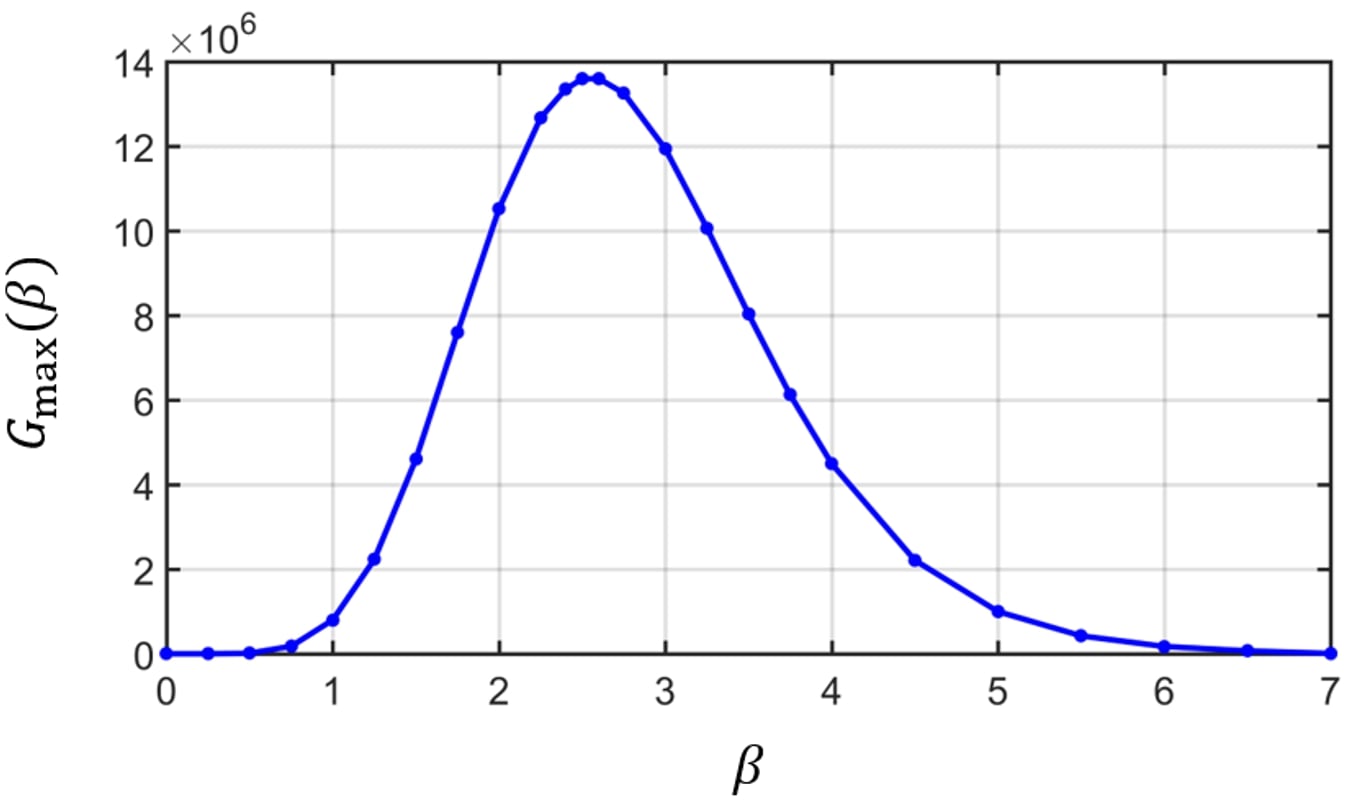}
	\caption{Maximum transient energy growth versus nondimensional spanwise wavenumber of a Mach $5.92$ SWBLI with an incident shock angle of $\theta_1=13\degree$ and a streamwise extent of $235\delta^*_\mathrm{in}$.}
	\label{beta_xsbli}
		% \vspace*{-0.25cm}
\end{figure}

 \tc{black}{We utilize the power iteration method with a random initial condition of unit energy to compute the optimal transient growth of the SWBLI in Figure \ref{base_xsbli}. Time interval for computing the optimal growth is set to roughly $2$ flow-through times, i.e., $(0, 450]$, where a flow-through time is defined as the time it takes a fluid particle that travels with the free-stream at Mach $5.92$ to traverse the entire streamwise length of the domain.} Similar to Figure \ref{beta_xsdbl}, we examine the dependence on the spanwise wavenumber $\beta$ of the largest transient growth in Figure \ref{beta_xsbli}. The peak transient growth of $1.36\times10^7$ at $\beta=2.6$ is roughly four orders of magnitude larger than the transient growth seen in Figure \ref{beta_xsdbl} for a streamwise domain length of $L_x=235$. This agrees with previous studies that often find that the formation of a recirculation bubble in a boundary-layer flow substantially increases the transient growth~\cite{Alizard,Cherubini}. The spanwise wavenumber also increases from $\beta=0.6$ for a nonparallel boundary layer to $\beta=2.6$ for an SWBLI. As the spanwise wavenumber deviates from the preferential value of $\beta=2.6$, the transient growth decreases by several orders of magnitude. 

\begin{figure}%[htb] % order of placement preference: here, top, bottom
	% \setstretch{1.}
	\includegraphics[width=12cm]{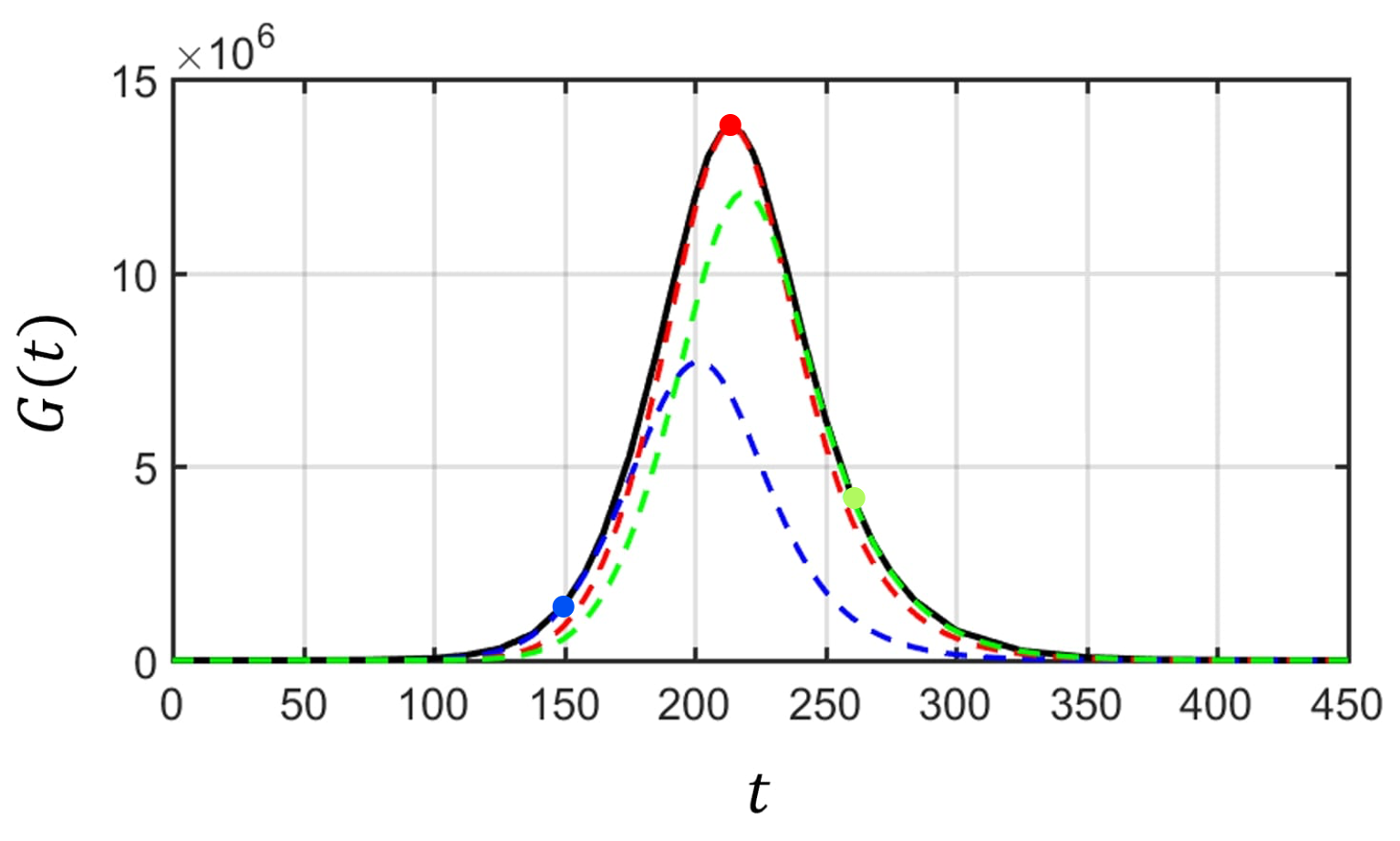}
	\caption{Transient growth envelope (solid black line) of a Mach 5.92 SWBLI with $\theta_1=13\degree$ and $\beta=2.6$. We show growth curves (dashed blue line, dashed red line, and dashed green line) that originate from three different initial conditions corresponding to the fixed time intervals $t=150$, $t=214$, and $t=260$. These transient growth curves touch the envelope at exactly one point (solid circle).}
	\label{grow_xsbli}
			\vspace*{-0.5cm}
\end{figure}

The transient growth envelope of the SWBLI in Figure \ref{base_xsbli} with $\beta=2.6$ is shown in Figure \ref{grow_xsbli}. Along with the temporal envelope, we show growth curves that originate from three different initial conditions pertaining to the fixed time intervals with $t=150$, $t=214$, and $t=260$. We note that in Figure \ref{grow_xsbli} each of the growth curves are tangential to the transient growth envelope at exactly one point and that the optimal transient growth of this SWBLI occurs at $t=214$. 

\begin{figure}%[H] % order of placement preference: here, top, bottom
	% \setstretch{1.}
	\centering
	\includegraphics[width=13cm]{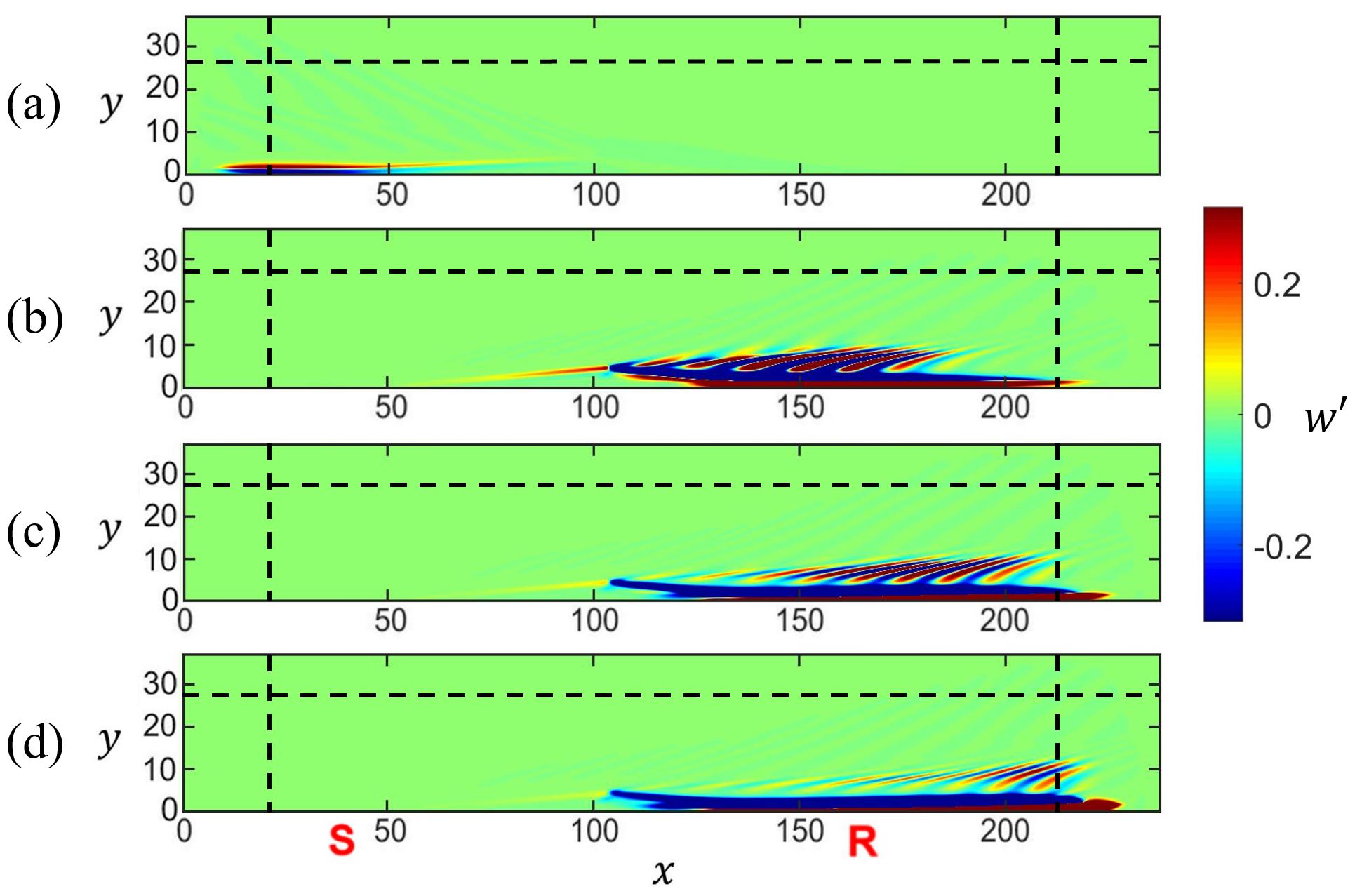}
	\caption{Evolution of the Mach 5.92 SWBLI with $\theta_1=13\degree$ and $\beta=2.6$ from the (a) optimal initial condition to its (d) final state at $t=214$. The two intermediary states occur at (b) $t = 150$ and (c) $t = 182$. Contour levels, representing the real part of the normalized spanwise velocity perturbation, are identical in each frame. The black dashes indicate the start of three different sponge layers.}
	\label{evo_xsbli}
\end{figure}

\tc{black}{Figure \ref{evo_xsbli} shows the normalized spanwise velocity perturbation component of the 2D optimal initial and final states in a Mach $5.92$ SWBLI with $\theta_1=13\degree$. We also examine the spatial structure of spanwise velocity perturbations at intermediate times $t=150$ and $t=182$ during amplification.} The evolution from the optimal initial condition corresponds to the transient growth curve with red dashes in Figure \ref{grow_xsbli}. \tc{black}{By $t=150$, barely any transient growth is observed but, after $t=150$,} there is a significant increase in the transient growth that persists until $t=214$. During the same time interval, in Figure \ref{evo_xsbli} we observe a large spanwise velocity perturbation that arises, at least in part, from the shear layer on top of the recirculation bubble and convects downstream until it hits the right sponge layer. \tc{black}{This observation is in agreement with previous studies which identified perturbations that arise from the shear layer and the decelerated zone in low-speed separated boundary-layer flows~\cite{Alizard,Cherubini,Akervik2}. When this large perturbation starts to decrease as it moves through the right sponge layer, the transient growth curve in Figure \ref{grow_xsbli} starts to decrease.}

\begin{figure}%[htb] % order of placement preference: here, top, bottom
	% \setstretch{1.}
	\includegraphics[width=11.4cm]{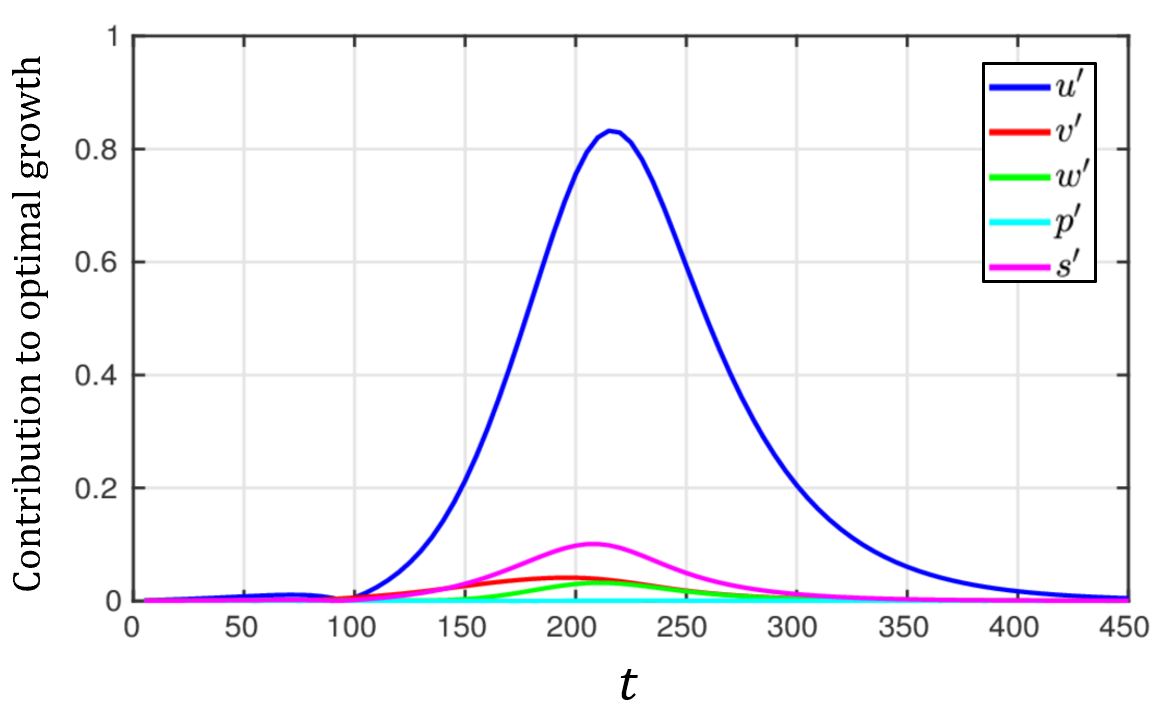}
	\caption{Contributions of the streamwise, wall-normal, and spanwise velocity perturbations as well as the entropy and pressure perturbations to the optimal transient growth of a Mach 5.92 SWBLI at $\theta_1=13\degree$ and $\beta=2.6$. Each curve that is associated with a perturbation variable has been normalized by the maximum growth at $t=214$.}
	\label{uvwps_contribute}
\end{figure}

\tc{black}{To determine the relative contribution of different perturbation components to the overall energy growth, in Figure \ref{uvwps_contribute} we evaluate the average magnitude of the velocity components as well as entropy and pressure as the optimal response evolves in time. We compute this average by integrating and normalizing magnitudes of flow perturbations in the domain at $t=214$.} The streamwise velocity perturbation has the most significant contribution to the optimal transient growth, which corresponds to the formation of streamwise elongated streaks downstream of the recirculation bubble. 

\begin{figure}%[htb] % order of placement preference: here, top, bottom
	% \setstretch{1.}
	\includegraphics[width=16cm]{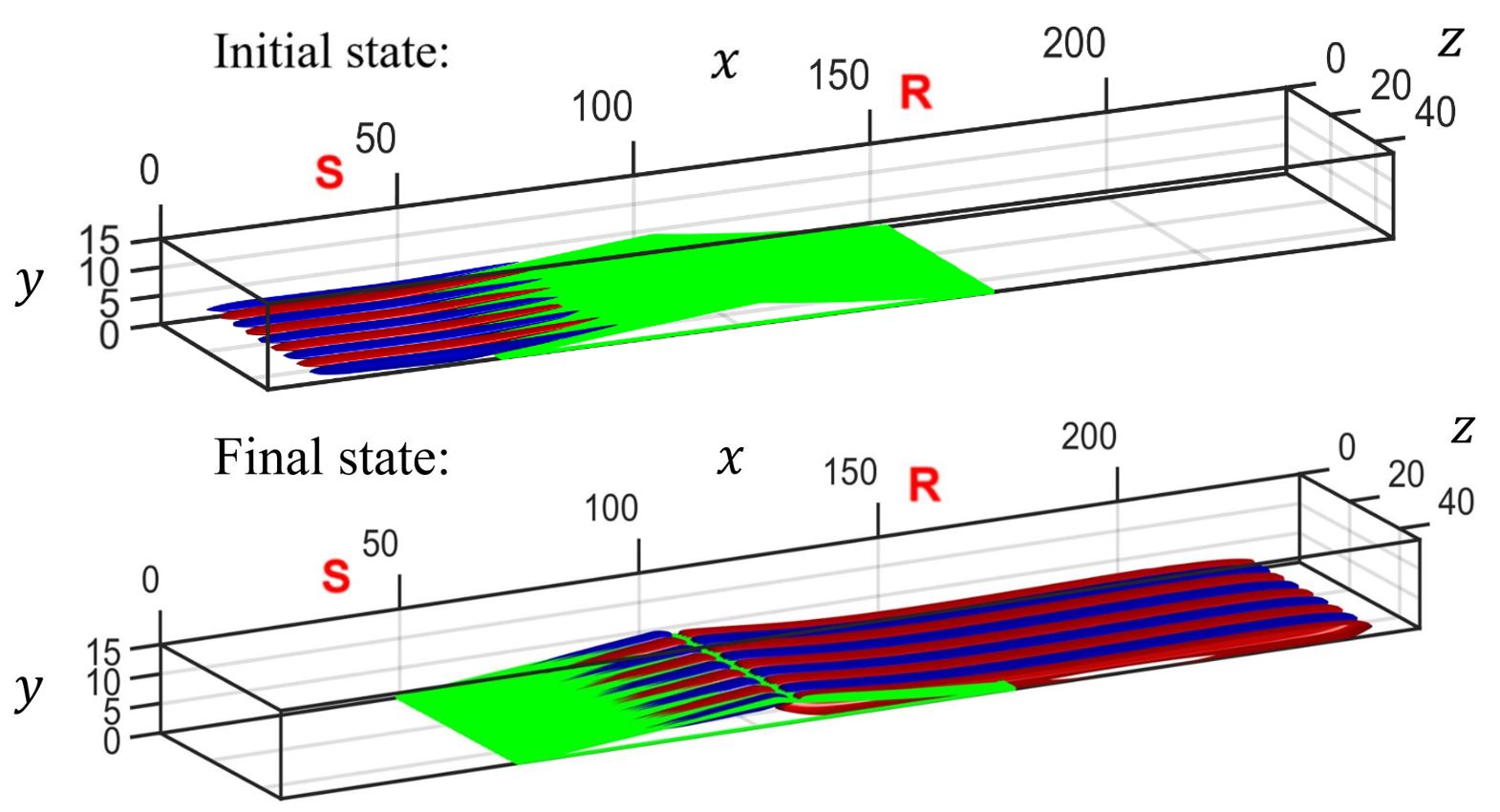}%%\vspace{0.4cm}
	\caption{Optimal initial and final states of a Mach 5.92 SWBLI with $\theta_1=13\degree$ and $\beta=2.6$. Isosurface contours, which represent the normalized streamwise velocity perturbation, are identical in each frame. The red and blue contours indicate positive and negative velocities, respectively, while the green contours indicate the size and position of the recirculation bubble.}
	\label{io_xsbli}
			\vspace*{-0.5cm}
\end{figure}

\tc{black}{Figure~\ref{io_xsbli} shows the streamwise velocity component of the 3D optimal initial and final states of a Mach $5.92$ SWBLI. The initial state consists of streamwise elongated structures that start near the inflow and extend past the separation point at $x\approx50$. As these structures evolve downstream, they result in the formation of the spanwise periodic streamwise streaks. To illustrate the relative location of these structures, we also plot the isosurface (in green) corresponding to the 2D recirculation zone. We note that the optimal transient response of this oblique SWBLI is very similar to the optimal frequency response observed in an SWBLI over a compression ramp that was identified via input-output analysis~\cite{Dwivedi2,Dwivedi3}.}

	\vspace*{-4ex}
\section{PHYSICAL MECHANISM OF THE OPTIMAL SWBLI RESPONSE}
\label{mech}

	\vspace*{-2ex}
\tc{black}{As demonstrated in Section~\ref{nonparallel}, the optimal spatio-temporal response in spatially-developing boundary layer consists of three-dimensional oblique flow structures. However, after we incorporate an incident oblique shock wave, streamwise streaks that experience significantly larger transient growth emerge. Herein, we investigate physical mechanisms responsible for large amplification by analyzing the spatio-temporal evolution of the optimal response and the dominant terms in the linearized inviscid transport equations.}

	\vspace*{-4ex}
\subsection{Wavepacket analysis}
\label{sec:wave}
\begin{figure}%[htb] % order of placement preference: here, top, bottom
	% \setstretch{1.}
	\includegraphics[width=16.5cm]{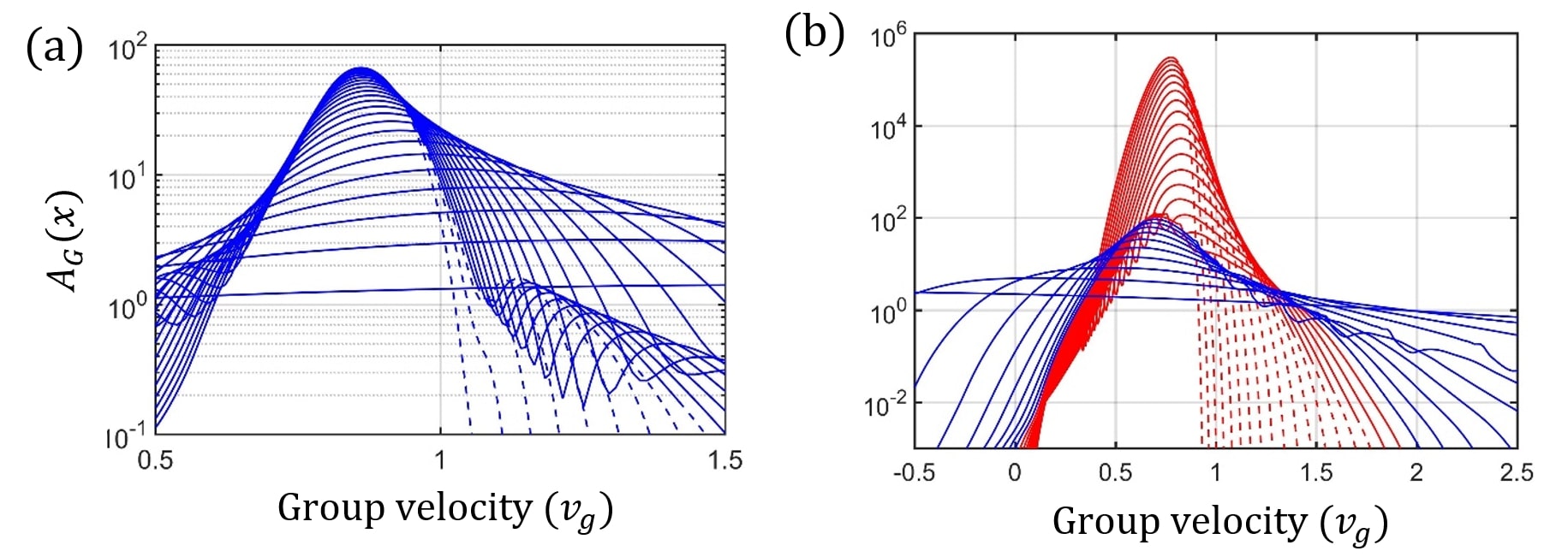}
	\caption{Wavepacket responses of the (a) spatially-developing boundary layer at $\beta=0.6$ and (b) oblique SWBLI at $\beta=2.6$. The dashed lines indicate the component of the wavepacket that resides within a sponge layer. For the oblique SWBLI at $\theta_1=13\degree$ in (b), the red contours represent times when the main peak of the wavepacket has crossed the apex of the recirculation bubble.}
	\label{group_velocity}
\end{figure}

	\vspace*{-2ex}
 \tc{black}{Figures \ref{io_xsdbl}, \ref{evo_xsbli}, and \ref{io_xsbli} show that the optimal spatio-temporal responses of both the spatially-developing boundary layer and the oblique SWBLI consist of wavepackets that amplify as they propagate downstream. Wavepackets occur in a variety of convectively-unstable flows including low-speed boundary layers~\cite{Akervik1} and high-speed jets~\cite{Nichols4}, and they have been linked to transient growth when viewed in the global framework~\cite{Chomaz}. To quantify wavepackets in our calculations, we follow the procedure described by Gallaire \& Chomaz~\cite{Gallaire}. First, we define the disturbance amplitude $A(x,t)$ to be the square root of the Chu energy integrated in the wall-normal direction at each streamwise location $x$ and time $t$. At $t=0$, the amplitude $A(x,0)$ peaks at $x = x_0$, which we take as the initial position of the wavepacket. We then consider the spatio-temporal response of the linearized flow equations along rays of constant group velocity $v_g = (x-x_0)/t$ resulting from this initial position. For the boundary layer with and without SWBLI, Figure~\ref{group_velocity} shows snapshots of the amplitude $A$ plotted against the group velocity $v_g$ at different instants of time throughout the optimal transient responses. For the boundary layer without SWBLI, Figure \ref{group_velocity}(a) reveals that amplification occurs for $v_g \in (0.7, 0.9)$. This means that an observer moving in a frame of reference moving to the right with velocity $v_g = 0.8$ will find that perturbations grow in time, thereby implying convective instability. For group velocities outside this range perturbations decay in time. The blue and red colors in this plot, respectively, indicate times before and after the wavepacket peak reaches the apex of the recirculation bubble. At early times, we observe amplification comparable to that in the entire boundary layer over the extent of the separating flow. Once the wavepacket reaches the bubble apex the amplification rate dramatically increases, suggesting the presence of a different physical mechanism for perturbation growth in the presence of SWBLI.}

 \tc{black}{As discussed in \cite{Gallaire,Delbende}, the amplification of wavepackets may be linked quantitatively to linear stability theory for locally parallel flows.  In a frame of reference moving with group velocity $v_g$, the temporal growth rate is}
\begin{equation}
	\sigma(v_g) 
	\; = \; 
	\lim\limits_{t \, \rightarrow \, \infty}\, \frac{\partial}{\partial t} \left( t^{1/2} A(x_0 + v_g t, t) \right) 
	\; \approx \; 
	\frac{ \ln \, (A(x_0 + v_g t_2, t_2)/A(x_0 + v_g t_1, t_1)}{t_2 \, - \, t_1} 
	\; + \;
	\frac{\ln \, (t_2/t_1)}{2(t_2 \, - \, t_1)},
\end{equation}
 \tc{black}{where the $t^{1/2}$-dependence accounts for the natural spread of the wavepacket. Here, $t_1$ and $t_2$ represent two time instants which are taken far enough apart to yield an accurate approximation~\cite{Delbende}. In the fixed frame, the growth rate and the streamwise wavenumber are determined by}
\begin{equation}
\label{omega_i}
	\begin{array}{rcl}
	\omega_i 
	& = & 
	\sigma(v_g) \; - \; v_g \, \dfrac{\sigma(v_g+\delta{x}/t_2) \, - \, \sigma(v_g)}{\delta{x}/t_2}, 
	\\[0.35cm] 
	\alpha(v_g) 
	& = & 
	\dfrac{\phi(x_o+v_gt_2+\delta{x},t_2) \, - \, \phi(x_o+v_gt_2,t_2)}{\delta{x}},
	\end{array}
\end{equation}
\tc{black}{where $\phi$ denotes the phase angle of the resulting wavepacket, and $\delta x$ represents a small change in the streamwise position. The second term on the right-hand-side of the expression for $\omega_i$ in Eq.~(\ref{omega_i}) accounts for spatial growth when converting between moving and fixed reference frames.  At the wavepacket peak, this term is zero and the growth rate equals the maximum temporal growth rate $\omega_{i,\text{max}}$ which is obtained by solving the stability problem for the linearization around locally parallel flow over all real wavenumbers. }

 \tc{black}{For the boundary layer without SWBLI, the wavepacket peak shown in Figure~\ref{group_velocity} gives $\omega_{i,\text{max}} = 0.0076$ at the streamwise wavenumber $\alpha = 0.11$ for the spanwise wavenumber $\beta = 0.6$.  Linear stability theory predicts $\omega_{i,\text{max}} =0.004 $ and $\alpha=0.15$, for the Mack first-mode instability at the same spanwise wavenumber. Qualitative agreement between these results indicates that the Mack first-mode instability plays a key role in the optimal transient response consisting of oblique waves observed in  the boundary layer without SWBLI. Similarly, for the wavepacket response in the presence of the oblique shock, we apply Eqs.~\eqref{omega_i} to the blue and the red curves in Figure~\ref{group_velocity}(b). The blue curves correspond to the time instances before the wavepacket reaches the apex of the recirculation bubble. From these curves, we obtain a maximum temporal growth rate of $\omega_{i,\text{max}} = 0.01$ with wavenumbers $\alpha = 0.11$ and $\beta = 2.6$. These initial perturbations are significantly less oblique than those obtained for the setup without the SWBLI. Linear theory predicts stability of the flow at this spanwise wavenumber and the observed growth rate $\omega_{i,\text{max}} = 0.01$ is not related to the Mack-first mode. The peak associated with the red curves in Figure~\ref{group_velocity}(b) for the time instants after the wavepacket crosses the recirculation bubble apex yields a growth rate of $\omega_{i,\text{max}} = 0.0768$, which is almost an order of magnitude larger than the growth rate of the upstream response. These structures correspond to the long streamwise streaks that form downstream and the streamwise wavenumber $\alpha = 0.029$ is much smaller than that observed upstream.} 
 
 		\vspace*{-5ex}
\subsection{Inviscid transport analysis}
	\label{sec.transport}

\begin{figure}
	% \setstretch{1.}
	\includegraphics[width=0.8\textwidth]{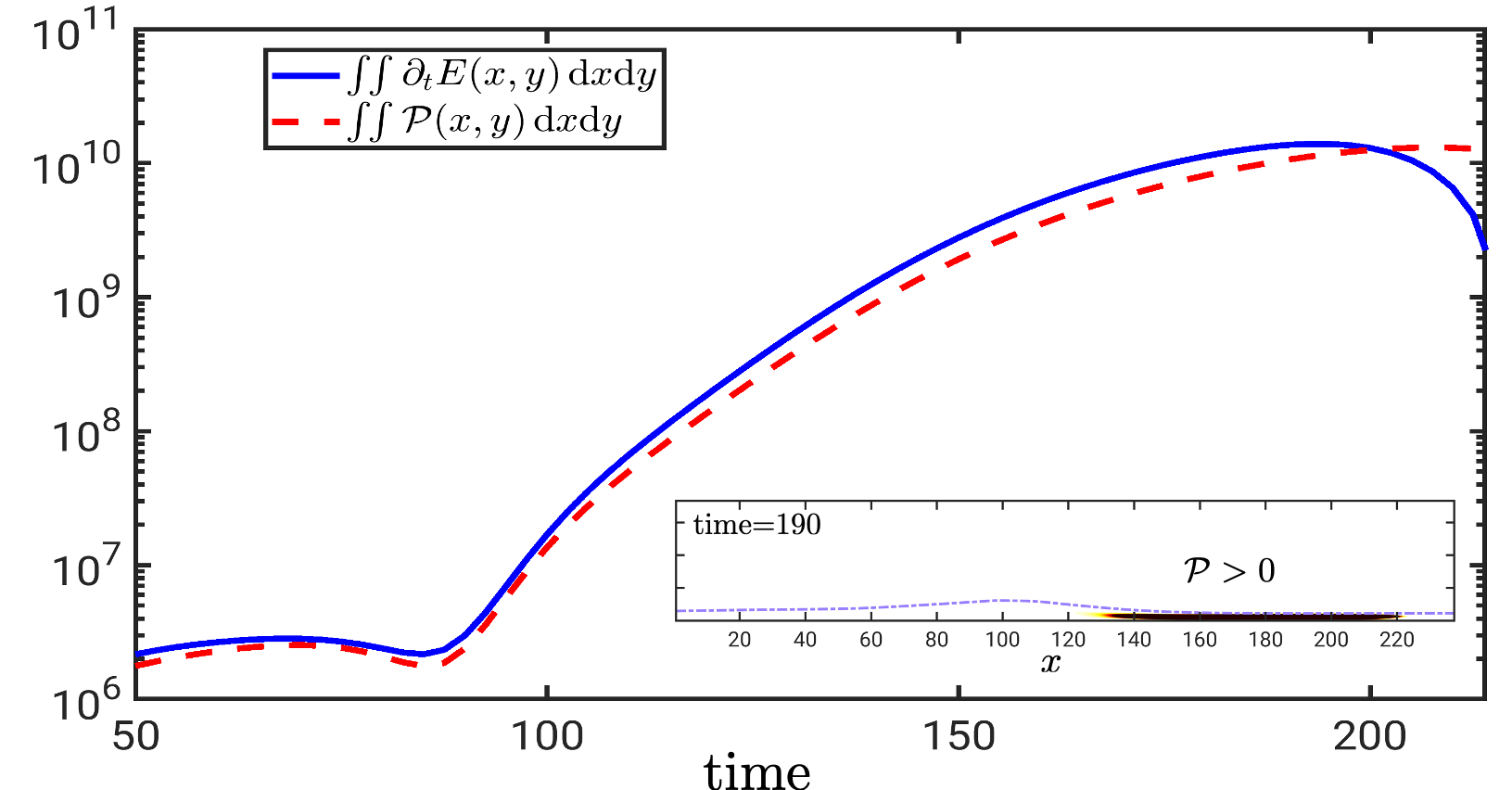}
	\caption{Comparison of the temporal change in the perturbation energy $\partial_t{E}$ (blue solid line) versus the production term $\mathcal{P}$ (red dashed line) integrated in space and plotted versus time for a Mach 5.92 SWBLI with $\theta_1=13\degree$ and $\beta=2.6$.  The inset shows the spatial region where $\mathcal{P}>0$ at $t = 190$.}
	\label{fig:PE_evol}
			\vspace*{-0.5cm}
\end{figure}

		\vspace*{-2ex}
\tc{black}{Our wavepacket analysis illustrates that the mechanism responsible for transient growth in SBLI is different from from the mechanism in a flat plate boundary layer.  To investigate this further, we examine the equation that governs the evolution of Chu's compressible energy $E$~\cite{Sidharth},} 
\begin{equation}
	\frac{\partial E}{\partial{t}} 
	\; = \; 
	\mathcal{P} \; + \; \mathcal{S} \; + \; \mathcal{V} \; - \; \mathcal{T}.
\label{ite}
\end{equation}
\tc{black}{Here, $\mathcal{P}$ represents the production of perturbations resulting from the base flow gradients, $\mathcal{S}$ denotes the source term that corresponds to the perturbation component of the inviscid material derivative, $\mathcal{V}$ accounts for the viscous dissipation, and $\mathcal{T}$ arises from advection of perturbations by the base-flow velocity. We evaluate the contribution of each term for the optimal response at $\beta=2.6$ as it amplifies in time (for $T\in[50,214]$). Figure~\ref{fig:PE_evol}  demonstrates that the production term $\mathcal{P}$ provides the dominant source of energy amplification. The inset in Figure~\ref{fig:PE_evol} shows that, at the time when energy achieves its largest value ($t = 190$), the production term $\mathcal{P}$ is active in the region near and after the reattachment location at $x=155$.}

\tc{black}{To investigate physical mechanisms which contribute to $\mathcal{P}$, we analyze  the  dominant  terms  in  the  corresponding linearized  inviscid  transport equations. As shown in Figure~\ref{uvwps_contribute}, the streamwise velocity component contributes the most to the kinetic energy. Motivated by this observation, we closely examine the spatio-temporal development of the streamwise momentum and  identify  amplification  mechanisms  that result from the interactions of flow perturbations with base flow gradients. After accounting for the dominant production terms, the transport of streamwise velocity is given by }
\begin{equation}
\frac{D u'}{D t}  
	\; \approx \;
	- \frac{1}{\bar{\rho}}\frac{\partial \bar{U}}{\partial x}{(\rho u)'} 
	\; - \; 
	\frac{1}{{\bar{\rho}}}\frac{\partial \bar{U}}{\partial y}{(\rho v)'}, 
\label{eq:transport}
\end{equation}
\tc{black}{where $D/Dt = {\partial}/{\partial t}\, + \, \bar{U}{\partial}/{\partial x} \,+\, \bar{V}{\partial}/{\partial y}$ quantifies the spatio-temporal evolution of $u'$ as it is advected by the base flow $(\bar{U},\bar{V})$. The first term on the right-hand-side results in the growth of $u'$ because of the streamwise deceleration of the base flow when ${\partial \bar{U}}/{\partial x} < 0$~\cite{Sidharth}, whereas the second term accounts for the growth that arises from the lift-up mechanism~\cite{Landahl1,Landahl2}. Multiplying Eq.~(\ref{eq:transport}) with $u'^{*}$ and integrating along the wall-normal and spanwise directions yields,}
\begin{equation}
\frac{1}{2} \, \Int_{0}^{y_0, \, z_{0}} \frac{D |u'|^2}{D t} \, \mathrm{d} y \, \mathrm{d} z 
	\; \approx \;
	- \Int_{0}^{y_0, \, z_{0}} \frac{1}{\bar{\rho}}\frac{\partial \bar{U}}{\partial x} \, {u'^*(\rho u)'} \, \mathrm{d} y \, \mathrm{d} z  
	\; - \; 
	\Int_{0}^{y_0, \, z_{0}}  \frac{1}{\bar{\rho}}\frac{\partial \bar{U}}{\partial y} \, {u'^*(\rho v)'} \, \mathrm{d} y \, \mathrm{d} z,
\label{eq:transport1}
\end{equation}
\tc{black}{where, $y_{0} = y_\text{max}$ is the wall-normal extent of the domain and $z_{0} = 2\pi/\beta$ (here, $\beta = 2.6$). We quantify the contribution of the streamwise velocity component to the energy of flow perturbations by examining the two terms on the right-hand-side of Eq.~\eqref{eq:transport1}.}

\begin{figure}%[htb] % order of placement preference: here, top, bottom
	% \setstretch{1.}
	\includegraphics[width=0.9\textwidth]{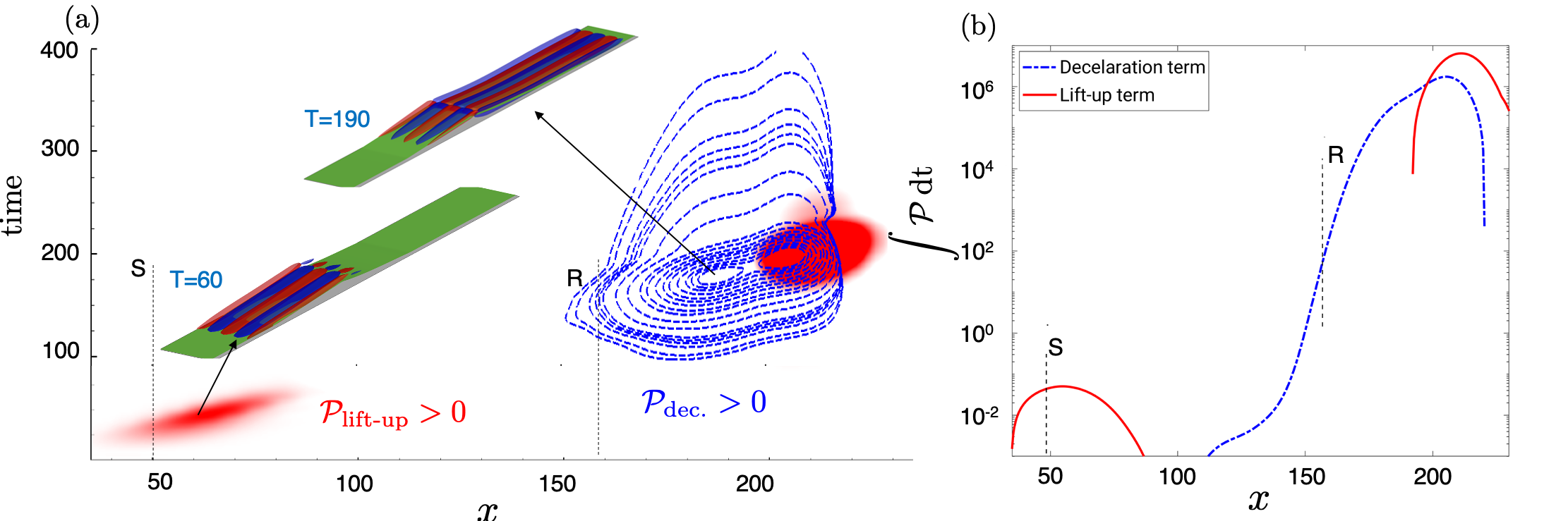}
	\caption{ \tc{black}{(a) Spatio-temporal evolution of production terms contributing to the growth of streamwise perturbation energy that arises from lift-up and streamwise deceleration. The inset shows the 3D iso-surfaces of the streamwise velocity along with separation streamline (in green). (b) Streamwise evolution of the time averaged contribution. The points S and R mark the location of separation and reattachment in the 2D steady flow. }  }
	\label{decel_shear}
			\vspace*{-0.25cm}
\end{figure}

\tc{black}{Figure~\ref{decel_shear}(a) shows the streamwise variation of the positive contribution that arises from the production terms as the flow perturbation evolve with time. Initially, the perturbations consist of streamwise vortices in the separating shear layer and they amplify via the lift-up mechanism. As perturbations develop in time and interact with the recirculation bubble the amplification mechanism changes and the streamwise deceleration term becomes dominant. To compare the spatial amplification resulting from these mechanisms, we also compute the time-averaged production terms. Figure~\ref{decel_shear}(b) illustrates that streamwise deceleration dominates the production of streamwise specific kinetic energy. This is in concert with a recent study of SWBLI over a compression ramp~\cite{Dwivedi3} which demonstrated that streamwise deceleration generates significant spatial amplification of streamwise streaks. Furthermore, inside the bubble, there is almost no contribution from the lift-up mechanism and its effects are only present near the separation point and significantly downstream of the reattachment point. }

\tc{black}{In addition to transient growth, our SWBLI configuration also supports a weak global instability that appears in the form of spanwise periodic streamwise streaks downstream of the recirculation bubble; also see~\cite{Hildebrand,Sidharth}. While the streaky structure of the global instability bears some similarity to the transient growth investigated in this paper, the global instability occurs for significantly lower spanwise wavenumbers ($\beta = 0.25$).  Also, the weak growth rate of the global instability ($\omega_{i}=2.5\times 10^{-3}$) induces a significantly smaller amplification over the time intervals for which the largest transient growth occurs. This difference in the most-amplified wavelength between global instability and transient growth analysis is a consequence of significant non-modal effects in high-speed boundary layers~\cite{Dwivedi3}. These effects make the SWBLI configuration extremely sensitive to external disturbances~\cite{Schmid2,jovARFM20} and, in the presence of experimental imperfections, they can trigger non-linear interactions, destabilize the reattached boundary layer, and induce \mbox{transition to turbulence.}}

\tc{black}{In summary, the wavepacket analysis of Section~\ref{sec:wave} shows that while the optimal response in the absence of SWBLI consists of oblique waves, the streamwise elongated streaks emerge in the presence of the shock. Furthermore, our analysis of the inviscid transport equation in Section~\ref{sec.transport} identifies physical mechanisms for transient growth and provides valuable insights into differences in the presence and absence of the shock. Even though local dispersion relations in both flows are similar before the separation location, in the absence of SWBLI most growth appears at the end of the domain. This growth arises from Mack's first mode instability and it is triggered by the oblique initial conditions that are tilted against the mean shear. In contrast, in the presence of the SBLI, amplification mostly arises from streamwise streaky perturbations as they convect to the reattachment location. In this case, the optimal initial condition takes the form of streamwise vortices. These are initially amplified by the lift-up mechanism and result in flow structures that can undergo considerable growth by streamwise compression.}

		% \vspace*{-4ex}
\section{CONCLUDING REMARKS}
\label{conclude}

		\vspace*{-2ex}
\tc{black}{We examined optimal transient growth in a high-speed compressible boundary layer interacting with an oblique shock wave.  To capture the effect of flow separation and recirculation caused by SWBLI, we utilize global linearized flow equations to identify the spatial structure of optimal initial perturbations and study the spatio-temporal evolution of the resulting response. For a hypersonic boundary layer without SWBLI, we verified that our approach correctly captures the dominant amplification that arises from the inviscid Orr mechanism, the lift-up effect, and the first mode instability. }

\tc{black}{To investigate the potential of shock wave/boundary-layer interactions to initiate transition, we examine transient growth of a hypersonic flat plate boundary layer with an oblique shock impinging on it. While the flow supports weak global instability, our analysis predicts large transient growth over short time periods. We demonstrate that the optimal initial condition is given by spanwise periodic vortices which commonly appear in the presence of leading-edge imperfections and distributed roughness in experiments. These initial perturbations results in the formation of streamwise streaks which are ubiquitous in experimental studies that involve SWBLI.}

\tc{black}{We also uncover physical mechanisms responsible for transient growth by quantifying the temporal evolution of dominant flow structures in various regions of the flow field. This is done by evaluating the contribution of the base-flow gradients to the production of streamwise kinetic energy in the inviscid transport equations. Our analysis demonstrates that streamwise streaks that emerge through the lift-up mechanism are significantly amplified by streamwise deceleration in the separation bubble near reattachment. Similar observations in other high-speed flow configurations with SWBLI, including compression ramps and double wedges, suggest that streamwise deceleration provides a robust physical mechanism for amplification of flow perturbations that solely arise in compressible separated flows. Our ongoing effort focuses on examining the influence of the Mach number as well as the initial angle of the incident shock and the angle after the incident shock interacts with the bow shock on transition mechanisms and dominant flow structures.}
   
\tc{black}{Transient growth analysis provides a useful framework for quantifying non-modal amplification mechanisms in complex high-speed flows even in the presence of global instabilities. This approach can provide important physical insights about the transition to turbulence, especially in experimental facilities where test times may be limited and where it is difficult to precisely control possible sources of external excitation.}

	\vspace*{-2ex}
\section*{ACKNOWLEDGMENTS}

	\vspace*{-2ex}
Financial support from the Office of Naval Research (ONR) under Awards N00014-15-1-2522 and N00014-17-1-2496 and from the Air Force Office of Scientific Research (AFOSR) under Award FA9550-18-1-0422 is gratefully acknowledged. The views and conclusions contained herein are those of the authors and should not be interpreted as representing the official policies or endorsements, either expressed or implied, of the ONR, the AFOSR, or the US Government.

	\newpage
%\bibliographystyle{apsrev4-1_custom}	
%\bibliography{bibliography/natePRF}

%merlin.mbs apsrev4-1.bst 2010-07-25 4.21a (PWD, AO, DPC) hacked
%Control: key (0)
%Control: author (72) initials jnrlst
%Control: editor formatted (1) identically to author
%Control: production of article title (1) required
%Control: page (0) single
%Control: year (1) truncated
%Control: production of eprint (0) enabled
\begin{thebibliography}{58}%
\makeatletter
\providecommand \@ifxundefined [1]{%
 \@ifx{#1\undefined}
}%
\providecommand \@ifnum [1]{%
 \ifnum #1\expandafter \@firstoftwo
 \else \expandafter \@secondoftwo
 \fi
}%
\providecommand \@ifx [1]{%
 \ifx #1\expandafter \@firstoftwo
 \else \expandafter \@secondoftwo
 \fi
}%
\providecommand \natexlab [1]{#1}%
\providecommand \enquote  [1]{``#1''}%
\providecommand \bibnamefont  [1]{#1}%
\providecommand \bibfnamefont [1]{#1}%
\providecommand \citenamefont [1]{#1}%
\providecommand \href@noop [0]{\@secondoftwo}%
\providecommand \href [0]{\begingroup \@sanitize@url \@href}%
\providecommand \@href[1]{\@@startlink{#1}\@@href}%
\providecommand \@@href[1]{\endgroup#1\@@endlink}%
\providecommand \@sanitize@url [0]{\catcode `\\12\catcode `\$12\catcode
  `\&12\catcode `\#12\catcode `\^12\catcode `\_12\catcode `\%12\relax}%
\providecommand \@@startlink[1]{}%
\providecommand \@@endlink[0]{}%
\providecommand \url  [0]{\begingroup\@sanitize@url \@url }%
\providecommand \@url [1]{\endgroup\@href {#1}{\urlprefix }}%
\providecommand \urlprefix  [0]{URL }%
\providecommand \Eprint [0]{\href }%
\providecommand \doibase [0]{http://dx.doi.org/}%
\providecommand \selectlanguage [0]{\@gobble}%
\providecommand \bibinfo  [0]{\@secondoftwo}%
\providecommand \bibfield  [0]{\@secondoftwo}%
\providecommand \translation [1]{[#1]}%
\providecommand \BibitemOpen [0]{}%
\providecommand \bibitemStop [0]{}%
\providecommand \bibitemNoStop [0]{.\EOS\space}%
\providecommand \EOS [0]{\spacefactor3000\relax}%
\providecommand \BibitemShut  [1]{\csname bibitem#1\endcsname}%
\let\auto@bib@innerbib\@empty
%</preamble>
\bibitem [{\citenamefont {Currao}\ \emph {et~al.}(2020)\citenamefont {Currao},
  \citenamefont {Choudhury}, \citenamefont {Gai}, \citenamefont {Neely},\ and\
  \citenamefont {Buttsworth}}]{Currao2019}%
  \BibitemOpen
  \bibfield  {author} {\bibinfo {author} {\bibfnamefont {G.~M.~D.}\
  \bibnamefont {Currao}}, \bibinfo {author} {\bibfnamefont {R.}~\bibnamefont
  {Choudhury}}, \bibinfo {author} {\bibfnamefont {S.~L.}\ \bibnamefont {Gai}},
  \bibinfo {author} {\bibfnamefont {A.~J.}\ \bibnamefont {Neely}}, \ and\
  \bibinfo {author} {\bibfnamefont {D.~R.}\ \bibnamefont {Buttsworth}},\
  }\bibfield  {title} {\enquote {\bibinfo {title} {Hypersonic transitional
  shock-wave--boundary-layer interaction on a flat plate},}\ }\href@noop {}
  {\bibfield  {journal} {\bibinfo  {journal} {AIAA J.}\ }\textbf {\bibinfo
  {volume} {58}},\ \bibinfo {pages} {814} (\bibinfo {year} {2020})}\BibitemShut
  {NoStop}%
\bibitem [{\citenamefont {Sandham}\ \emph {et~al.}(2014)\citenamefont
  {Sandham}, \citenamefont {Schulein}, \citenamefont {Wagner}, \citenamefont
  {Willems},\ and\ \citenamefont {Steelant}}]{Sandham}%
  \BibitemOpen
  \bibfield  {author} {\bibinfo {author} {\bibfnamefont {N.~D.}\ \bibnamefont
  {Sandham}}, \bibinfo {author} {\bibfnamefont {E.}~\bibnamefont {Schulein}},
  \bibinfo {author} {\bibfnamefont {A.}~\bibnamefont {Wagner}}, \bibinfo
  {author} {\bibfnamefont {S.}~\bibnamefont {Willems}}, \ and\ \bibinfo
  {author} {\bibfnamefont {J.}~\bibnamefont {Steelant}},\ }\bibfield  {title}
  {\enquote {\bibinfo {title} {Transitional shock-wave/boundary-layer
  interactions in hypersonic flow},}\ }\href@noop {} {\bibfield  {journal}
  {\bibinfo  {journal} {J. Fluid Mech.}\ }\textbf {\bibinfo {volume} {752}},\
  \bibinfo {pages} {349} (\bibinfo {year} {2014})}\BibitemShut {NoStop}%
\bibitem [{\citenamefont {Pagella}\ \emph {et~al.}(2002)\citenamefont
  {Pagella}, \citenamefont {Rist},\ and\ \citenamefont {Wagner}}]{Pagella1}%
  \BibitemOpen
  \bibfield  {author} {\bibinfo {author} {\bibfnamefont {A.}~\bibnamefont
  {Pagella}}, \bibinfo {author} {\bibfnamefont {U.}~\bibnamefont {Rist}}, \
  and\ \bibinfo {author} {\bibfnamefont {S.}~\bibnamefont {Wagner}},\
  }\bibfield  {title} {\enquote {\bibinfo {title} {Numerical investigations of
  small-amplitude disturbances in a boundary layer with impinging shock wave at
  ${M}\!a = 4.8$},}\ }\href@noop {} {\bibfield  {journal} {\bibinfo  {journal}
  {Phys. Fluids}\ }\textbf {\bibinfo {volume} {14}},\ \bibinfo {pages} {2088}
  (\bibinfo {year} {2002})}\BibitemShut {NoStop}%
\bibitem [{\citenamefont {Chuvakhov}\ and\ \citenamefont
  {Radchenko}(2020)}]{Chuvakhov2020}%
  \BibitemOpen
  \bibfield  {author} {\bibinfo {author} {\bibfnamefont {P.~V.}\ \bibnamefont
  {Chuvakhov}}\ and\ \bibinfo {author} {\bibfnamefont {V.~N.}\ \bibnamefont
  {Radchenko}},\ }\bibfield  {title} {\enquote {\bibinfo {title} {Effect of
  {G}{\"o}rtler-like vortices of various intensity on heat transfer in
  supersonic compression corner flows},}\ }\href@noop {} {\bibfield  {journal}
  {\bibinfo  {journal} {Int. J. of Heat and Mass Trans.}\ }\textbf {\bibinfo
  {volume} {150}},\ \bibinfo {pages} {119310 (12 pages)} (\bibinfo {year}
  {2020})}\BibitemShut {NoStop}%
\bibitem [{\citenamefont {Cao}\ \emph {et~al.}(2019)\citenamefont {Cao},
  \citenamefont {Klioutchnikov},\ and\ \citenamefont
  {Olivier}}]{CaoKliHer2019}%
  \BibitemOpen
  \bibfield  {author} {\bibinfo {author} {\bibfnamefont {S.}~\bibnamefont
  {Cao}}, \bibinfo {author} {\bibfnamefont {I.}~\bibnamefont {Klioutchnikov}},
  \ and\ \bibinfo {author} {\bibfnamefont {H.}~\bibnamefont {Olivier}},\
  }\bibfield  {title} {\enquote {\bibinfo {title} {G{\"o}rtler vortices in
  hypersonic flow on compression ramps},}\ }\href@noop {} {\bibfield  {journal}
  {\bibinfo  {journal} {AIAA J.}\ }\textbf {\bibinfo {volume} {57}},\ \bibinfo
  {pages} {3874} (\bibinfo {year} {2019})}\BibitemShut {NoStop}%
\bibitem [{\citenamefont {Navarro-Martinez}\ and\ \citenamefont
  {Tutty}(2005)}]{navarro2005}%
  \BibitemOpen
  \bibfield  {author} {\bibinfo {author} {\bibfnamefont {S.}~\bibnamefont
  {Navarro-Martinez}}\ and\ \bibinfo {author} {\bibfnamefont {O.~R.}\
  \bibnamefont {Tutty}},\ }\bibfield  {title} {\enquote {\bibinfo {title}
  {Numerical simulation of {G}{\"o}rtler vortices in hypersonic compression
  ramps},}\ }\href@noop {} {\bibfield  {journal} {\bibinfo  {journal} {Comp. \&
  Fluids}\ }\textbf {\bibinfo {volume} {34}},\ \bibinfo {pages} {225} (\bibinfo
  {year} {2005})}\BibitemShut {NoStop}%
\bibitem [{\citenamefont {Priebe}\ \emph {et~al.}(2016)\citenamefont {Priebe},
  \citenamefont {Tu}, \citenamefont {Rowley},\ and\ \citenamefont
  {Mart{\'\i}n}}]{priebe2016low}%
  \BibitemOpen
  \bibfield  {author} {\bibinfo {author} {\bibfnamefont {S.}~\bibnamefont
  {Priebe}}, \bibinfo {author} {\bibfnamefont {J.~H.}\ \bibnamefont {Tu}},
  \bibinfo {author} {\bibfnamefont {C.~W.}\ \bibnamefont {Rowley}}, \ and\
  \bibinfo {author} {\bibfnamefont {M.~P.}\ \bibnamefont {Mart{\'\i}n}},\
  }\bibfield  {title} {\enquote {\bibinfo {title} {Low-frequency dynamics in a
  shock-induced separated flow},}\ }\href@noop {} {\bibfield  {journal}
  {\bibinfo  {journal} {J. of Fluid Mech.}\ }\textbf {\bibinfo {volume}
  {807}},\ \bibinfo {pages} {441} (\bibinfo {year} {2016})}\BibitemShut
  {NoStop}%
\bibitem [{\citenamefont {Zhuang}\ \emph {et~al.}(2017)\citenamefont {Zhuang},
  \citenamefont {Tan}, \citenamefont {Liu}, \citenamefont {Zhang},\ and\
  \citenamefont {Ling}}]{Zhuang2017}%
  \BibitemOpen
  \bibfield  {author} {\bibinfo {author} {\bibfnamefont {Y.}~\bibnamefont
  {Zhuang}}, \bibinfo {author} {\bibfnamefont {H.~J.}\ \bibnamefont {Tan}},
  \bibinfo {author} {\bibfnamefont {Y.~Z.}\ \bibnamefont {Liu}}, \bibinfo
  {author} {\bibfnamefont {Y.~C.}\ \bibnamefont {Zhang}}, \ and\ \bibinfo
  {author} {\bibfnamefont {Y.}~\bibnamefont {Ling}},\ }\bibfield  {title}
  {\enquote {\bibinfo {title} {High resolution visualization of
  {G}{\"o}rtler-like vortices in supersonic compression ramp flow},}\
  }\href@noop {} {\bibfield  {journal} {\bibinfo  {journal} {J. of Vis.}\
  }\textbf {\bibinfo {volume} {20}},\ \bibinfo {pages} {505} (\bibinfo {year}
  {2017})}\BibitemShut {NoStop}%
\bibitem [{\citenamefont {Theofilis}\ \emph {et~al.}(2000)\citenamefont
  {Theofilis}, \citenamefont {Hein},\ and\ \citenamefont
  {Dallmann}}]{Theofilis}%
  \BibitemOpen
  \bibfield  {author} {\bibinfo {author} {\bibfnamefont {V.}~\bibnamefont
  {Theofilis}}, \bibinfo {author} {\bibfnamefont {S.}~\bibnamefont {Hein}}, \
  and\ \bibinfo {author} {\bibfnamefont {U.}~\bibnamefont {Dallmann}},\
  }\bibfield  {title} {\enquote {\bibinfo {title} {On the origins of
  unsteadiness and three-dimensionality in a laminar separation bubble},}\
  }\href@noop {} {\bibfield  {journal} {\bibinfo  {journal} {Philos. Trans.
  Roy. Soc. A}\ }\textbf {\bibinfo {volume} {358}},\ \bibinfo {pages} {3229}
  (\bibinfo {year} {2000})}\BibitemShut {NoStop}%
\bibitem [{\citenamefont {Theofilis}(2011)}]{TheofilisARFM}%
  \BibitemOpen
  \bibfield  {author} {\bibinfo {author} {\bibfnamefont {V.}~\bibnamefont
  {Theofilis}},\ }\bibfield  {title} {\enquote {\bibinfo {title} {Global linear
  instability},}\ }\href@noop {} {\bibfield  {journal} {\bibinfo  {journal}
  {Annu. Rev. of Fluid Mech.}\ }\textbf {\bibinfo {volume} {43}},\ \bibinfo
  {pages} {319} (\bibinfo {year} {2011})}\BibitemShut {NoStop}%
\bibitem [{\citenamefont {Hildebrand}\ \emph
  {et~al.}(2018{\natexlab{a}})\citenamefont {Hildebrand}, \citenamefont
  {Dwivedi}, \citenamefont {Nichols}, \citenamefont {Jovanovi\'c},\ and\
  \citenamefont {Candler}}]{Hildebrand}%
  \BibitemOpen
  \bibfield  {author} {\bibinfo {author} {\bibfnamefont {N.}~\bibnamefont
  {Hildebrand}}, \bibinfo {author} {\bibfnamefont {A.}~\bibnamefont {Dwivedi}},
  \bibinfo {author} {\bibfnamefont {J.~W.}\ \bibnamefont {Nichols}}, \bibinfo
  {author} {\bibfnamefont {M.~R.}\ \bibnamefont {Jovanovi\'c}}, \ and\ \bibinfo
  {author} {\bibfnamefont {G.~V.}\ \bibnamefont {Candler}},\ }\bibfield
  {title} {\enquote {\bibinfo {title} {Simulation and stability analysis of
  oblique shock wave/boundary layer interactions at {M}ach $5.92$},}\
  }\href@noop {} {\bibfield  {journal} {\bibinfo  {journal} {Phys. Rev.
  Fluids}\ }\textbf {\bibinfo {volume} {3}},\ \bibinfo {pages} {013906}
  (\bibinfo {year} {2018}{\natexlab{a}})}\BibitemShut {NoStop}%
\bibitem [{\citenamefont {Robinet}(2007)}]{Robinet}%
  \BibitemOpen
  \bibfield  {author} {\bibinfo {author} {\bibfnamefont {J.-C.}\ \bibnamefont
  {Robinet}},\ }\bibfield  {title} {\enquote {\bibinfo {title} {Bifurcations in
  shock-wave/laminar-boundary-layer interaction: global instability
  approach},}\ }\href@noop {} {\bibfield  {journal} {\bibinfo  {journal} {J.
  Fluid Mech.}\ }\textbf {\bibinfo {volume} {579}},\ \bibinfo {pages} {85}
  (\bibinfo {year} {2007})}\BibitemShut {NoStop}%
\bibitem [{\citenamefont {Sidharth}\ \emph {et~al.}(2018)\citenamefont
  {Sidharth}, \citenamefont {Dwivedi}, \citenamefont {Candler},\ and\
  \citenamefont {Nichols}}]{Sidharth}%
  \BibitemOpen
  \bibfield  {author} {\bibinfo {author} {\bibfnamefont {G.~S.}\ \bibnamefont
  {Sidharth}}, \bibinfo {author} {\bibfnamefont {A.}~\bibnamefont {Dwivedi}},
  \bibinfo {author} {\bibfnamefont {G.~V.}\ \bibnamefont {Candler}}, \ and\
  \bibinfo {author} {\bibfnamefont {J.~W.}\ \bibnamefont {Nichols}},\
  }\bibfield  {title} {\enquote {\bibinfo {title} {Onset of three
  dimensionality in supersonic flow over a slender double wedge},}\ }\href@noop
  {} {\bibfield  {journal} {\bibinfo  {journal} {Phys. Rev. Fluids}\ }\textbf
  {\bibinfo {volume} {3}},\ \bibinfo {pages} {093901 (29 pages)} (\bibinfo
  {year} {2018})}\BibitemShut {NoStop}%
\bibitem [{\citenamefont {Chuvakhov}\ \emph {et~al.}(2017)\citenamefont
  {Chuvakhov}, \citenamefont {Borovoy}, \citenamefont {Egorov}, \citenamefont
  {Radchenko}, \citenamefont {Olivier},\ and\ \citenamefont
  {Roghelia}}]{Chuvakhov2017}%
  \BibitemOpen
  \bibfield  {author} {\bibinfo {author} {\bibfnamefont {P.~V.}\ \bibnamefont
  {Chuvakhov}}, \bibinfo {author} {\bibfnamefont {V.~Y.}\ \bibnamefont
  {Borovoy}}, \bibinfo {author} {\bibfnamefont {I.~V.}\ \bibnamefont {Egorov}},
  \bibinfo {author} {\bibfnamefont {V.~N.}\ \bibnamefont {Radchenko}}, \bibinfo
  {author} {\bibfnamefont {H.}~\bibnamefont {Olivier}}, \ and\ \bibinfo
  {author} {\bibfnamefont {A.}~\bibnamefont {Roghelia}},\ }\bibfield  {title}
  {\enquote {\bibinfo {title} {Effect of small bluntness on formation of
  {G}{\"o}rtler vortices in a supersonic compression corner flow},}\
  }\href@noop {} {\bibfield  {journal} {\bibinfo  {journal} {J. Appl. Mech.
  Tech. Phys.}\ }\textbf {\bibinfo {volume} {58}},\ \bibinfo {pages} {975}
  (\bibinfo {year} {2017})}\BibitemShut {NoStop}%
\bibitem [{\citenamefont {Sidharth}\ \emph {et~al.}(2017)\citenamefont
  {Sidharth}, \citenamefont {Dwivedi}, \citenamefont {Candler},\ and\
  \citenamefont {Nichols}}]{gs2017global}%
  \BibitemOpen
  \bibfield  {author} {\bibinfo {author} {\bibfnamefont {G.~S.}\ \bibnamefont
  {Sidharth}}, \bibinfo {author} {\bibfnamefont {A.}~\bibnamefont {Dwivedi}},
  \bibinfo {author} {\bibfnamefont {G.~V.}\ \bibnamefont {Candler}}, \ and\
  \bibinfo {author} {\bibfnamefont {J.~W.}\ \bibnamefont {Nichols}},\
  }\bibfield  {title} {\enquote {\bibinfo {title} {Global linear stability
  analysis of high speed flows on compression ramps},}\ }in\ \href@noop {}
  {\emph {\bibinfo {booktitle} {Proceedings of the 47th AIAA Fluid Dynamics
  Conference}}}\ (\bibinfo {address} {Denver, CO},\ \bibinfo {year} {2017})\
  p.\ \bibinfo {pages} {3455 (14 pages)}\BibitemShut {NoStop}%
\bibitem [{\citenamefont {Butler}\ and\ \citenamefont
  {Farrell}(1992)}]{Butler}%
  \BibitemOpen
  \bibfield  {author} {\bibinfo {author} {\bibfnamefont {K.~M.}\ \bibnamefont
  {Butler}}\ and\ \bibinfo {author} {\bibfnamefont {B.~F.}\ \bibnamefont
  {Farrell}},\ }\bibfield  {title} {\enquote {\bibinfo {title}
  {Three-dimensional optimal perturbations in viscous shear flow},}\
  }\href@noop {} {\bibfield  {journal} {\bibinfo  {journal} {Phys. Fluids}\
  }\textbf {\bibinfo {volume} {4}},\ \bibinfo {pages} {1637} (\bibinfo {year}
  {1992})}\BibitemShut {NoStop}%
\bibitem [{\citenamefont {Andersson}\ \emph {et~al.}(1999)\citenamefont
  {Andersson}, \citenamefont {Berggren},\ and\ \citenamefont
  {Henningson}}]{Andersson}%
  \BibitemOpen
  \bibfield  {author} {\bibinfo {author} {\bibfnamefont {P.}~\bibnamefont
  {Andersson}}, \bibinfo {author} {\bibfnamefont {M.}~\bibnamefont {Berggren}},
  \ and\ \bibinfo {author} {\bibfnamefont {D.~S.}\ \bibnamefont {Henningson}},\
  }\bibfield  {title} {\enquote {\bibinfo {title} {Optimal disturbances and
  bypass transition in boundary layers},}\ }\href@noop {} {\bibfield  {journal}
  {\bibinfo  {journal} {Phys. Fluids}\ }\textbf {\bibinfo {volume} {11}},\
  \bibinfo {pages} {134} (\bibinfo {year} {1999})}\BibitemShut {NoStop}%
\bibitem [{\citenamefont {Jovanovi\'c}\ and\ \citenamefont
  {Bamieh}(2005)}]{Jovanovic}%
  \BibitemOpen
  \bibfield  {author} {\bibinfo {author} {\bibfnamefont {M.~R.}\ \bibnamefont
  {Jovanovi\'c}}\ and\ \bibinfo {author} {\bibfnamefont {B.}~\bibnamefont
  {Bamieh}},\ }\bibfield  {title} {\enquote {\bibinfo {title} {Componentwise
  energy amplification in channel flows},}\ }\href@noop {} {\bibfield
  {journal} {\bibinfo  {journal} {J. Fluid Mech.}\ }\textbf {\bibinfo {volume}
  {534}},\ \bibinfo {pages} {145} (\bibinfo {year} {2005})}\BibitemShut
  {NoStop}%
\bibitem [{\citenamefont {Schmid}\ and\ \citenamefont
  {Henningson}(2001)}]{Schmid1}%
  \BibitemOpen
  \bibfield  {author} {\bibinfo {author} {\bibfnamefont {P.~J.}\ \bibnamefont
  {Schmid}}\ and\ \bibinfo {author} {\bibfnamefont {D.~S.}\ \bibnamefont
  {Henningson}},\ }\href@noop {} {\emph {\bibinfo {title} {Stability and
  transition in shear flows}}}\ (\bibinfo  {publisher} {Springer-Verlag},\
  \bibinfo {address} {New York},\ \bibinfo {year} {2001})\BibitemShut {NoStop}%
\bibitem [{\citenamefont {Jovanovi\'c}(2004)}]{mj-phd04}%
  \BibitemOpen
  \bibfield  {author} {\bibinfo {author} {\bibfnamefont {M.~R.}\ \bibnamefont
  {Jovanovi\'c}},\ }\emph {\bibinfo {title} {Modeling, analysis, and control of
  spatially distributed systems}},\ \href@noop {} {Ph.D. thesis},\ \bibinfo
  {school} {University of California, Santa Barbara} (\bibinfo {year}
  {2004})\BibitemShut {NoStop}%
\bibitem [{\citenamefont {Schmid}(2007)}]{Schmid2}%
  \BibitemOpen
  \bibfield  {author} {\bibinfo {author} {\bibfnamefont {P.~J.}\ \bibnamefont
  {Schmid}},\ }\bibfield  {title} {\enquote {\bibinfo {title} {Nonmodal
  stability theory},}\ }\href@noop {} {\bibfield  {journal} {\bibinfo
  {journal} {Annu. Rev. Fluid Mech.}\ }\textbf {\bibinfo {volume} {39}},\
  \bibinfo {pages} {129} (\bibinfo {year} {2007})}\BibitemShut {NoStop}%
\bibitem [{\citenamefont {Jovanovi\'c}(2020)}]{jovARFM20}%
  \BibitemOpen
  \bibfield  {author} {\bibinfo {author} {\bibfnamefont {M.~R.}\ \bibnamefont
  {Jovanovi\'c}},\ }\bibfield  {title} {\enquote {\bibinfo {title} {From bypass
  transition to flow control and data-driven turbulence modeling: {A}n
  input-output viewpoint},}\ }\href@noop {} {\bibfield  {journal} {\bibinfo
  {journal} {Annu. Rev. Fluid Mech.}\ } (\bibinfo {year} {2020})},\ \bibinfo
  {note} {submitted; also arXiv:2003.10104}\BibitemShut {NoStop}%
\bibitem [{\citenamefont {Dwivedi}\ \emph {et~al.}(2019)\citenamefont
  {Dwivedi}, \citenamefont {Sidharth}, \citenamefont {Candler}, \citenamefont
  {Nichols},\ and\ \citenamefont {Jovanovi\'c}}]{Dwivedi3}%
  \BibitemOpen
  \bibfield  {author} {\bibinfo {author} {\bibfnamefont {A.}~\bibnamefont
  {Dwivedi}}, \bibinfo {author} {\bibfnamefont {G.~S.}\ \bibnamefont
  {Sidharth}}, \bibinfo {author} {\bibfnamefont {G.~V.}\ \bibnamefont
  {Candler}}, \bibinfo {author} {\bibfnamefont {J.~W.}\ \bibnamefont
  {Nichols}}, \ and\ \bibinfo {author} {\bibfnamefont {M.~R.}\ \bibnamefont
  {Jovanovi\'c}},\ }\bibfield  {title} {\enquote {\bibinfo {title}
  {Reattachment streaks in hypersonic compression ramp flow: an input-output
  analysis},}\ }\href@noop {} {\bibfield  {journal} {\bibinfo  {journal} {J.
  Fluid Mech.}\ }\textbf {\bibinfo {volume} {880}},\ \bibinfo {pages} {113}
  (\bibinfo {year} {2019})}\BibitemShut {NoStop}%
\bibitem [{\citenamefont {Dwivedi}\ \emph {et~al.}(2018)\citenamefont
  {Dwivedi}, \citenamefont {Sidharth}, \citenamefont {Candler}, \citenamefont
  {Nichols},\ and\ \citenamefont {Jovanovi\'c}}]{Dwivedi2}%
  \BibitemOpen
  \bibfield  {author} {\bibinfo {author} {\bibfnamefont {A.}~\bibnamefont
  {Dwivedi}}, \bibinfo {author} {\bibfnamefont {G.~S.}\ \bibnamefont
  {Sidharth}}, \bibinfo {author} {\bibfnamefont {G.~V.}\ \bibnamefont
  {Candler}}, \bibinfo {author} {\bibfnamefont {J.~W.}\ \bibnamefont
  {Nichols}}, \ and\ \bibinfo {author} {\bibfnamefont {M.~R.}\ \bibnamefont
  {Jovanovi\'c}},\ }\bibfield  {title} {\enquote {\bibinfo {title}
  {Input-output analysis of shock boundary layer interaction},}\ }in\
  \href@noop {} {\emph {\bibinfo {booktitle} {Proceedings of the 48th AIAA
  Fluid Dynamics Conference}}}\ (\bibinfo {address} {Atlanta, GA},\ \bibinfo
  {year} {2018})\ p.\ \bibinfo {pages} {3220 (17 pages)}\BibitemShut {NoStop}%
\bibitem [{\citenamefont {Hanifi}\ \emph {et~al.}(1996)\citenamefont {Hanifi},
  \citenamefont {Schmid},\ and\ \citenamefont {Henningson}}]{Hanifi}%
  \BibitemOpen
  \bibfield  {author} {\bibinfo {author} {\bibfnamefont {A.}~\bibnamefont
  {Hanifi}}, \bibinfo {author} {\bibfnamefont {P.~J.}\ \bibnamefont {Schmid}},
  \ and\ \bibinfo {author} {\bibfnamefont {D.~S.}\ \bibnamefont {Henningson}},\
  }\bibfield  {title} {\enquote {\bibinfo {title} {Transient growth in
  compressible boundary layer flow},}\ }\href@noop {} {\bibfield  {journal}
  {\bibinfo  {journal} {Phys. Fluids}\ }\textbf {\bibinfo {volume} {8}},\
  \bibinfo {pages} {826} (\bibinfo {year} {1996})}\BibitemShut {NoStop}%
\bibitem [{\citenamefont {Bitter}\ and\ \citenamefont
  {Shepherd}(2015)}]{Bitter1}%
  \BibitemOpen
  \bibfield  {author} {\bibinfo {author} {\bibfnamefont {N.~P.}\ \bibnamefont
  {Bitter}}\ and\ \bibinfo {author} {\bibfnamefont {J.~E.}\ \bibnamefont
  {Shepherd}},\ }\bibfield  {title} {\enquote {\bibinfo {title} {Stability of
  highly cooled hypervelocity boundary layers},}\ }\href@noop {} {\bibfield
  {journal} {\bibinfo  {journal} {J. of Fluid Mech.}\ }\textbf {\bibinfo
  {volume} {778}},\ \bibinfo {pages} {586} (\bibinfo {year}
  {2015})}\BibitemShut {NoStop}%
\bibitem [{\citenamefont {Bitter}(2015)}]{Bitter2}%
  \BibitemOpen
  \bibfield  {author} {\bibinfo {author} {\bibfnamefont {N.~P.}\ \bibnamefont
  {Bitter}},\ }\emph {\bibinfo {title} {Stability of hypervelocity boundary
  layers}},\ \href@noop {} {Ph.D. thesis},\ \bibinfo  {school} {California
  Institute of Technology} (\bibinfo {year} {2015})\BibitemShut {NoStop}%
\bibitem [{\citenamefont {Luchini}(2000)}]{Luchini2}%
  \BibitemOpen
  \bibfield  {author} {\bibinfo {author} {\bibfnamefont {P.}~\bibnamefont
  {Luchini}},\ }\bibfield  {title} {\enquote {\bibinfo {title}
  {Reynolds-number-independent instability of the boundary layer over a flat
  surface: optimal perturbations},}\ }\href@noop {} {\bibfield  {journal}
  {\bibinfo  {journal} {J. Fluid Mech.}\ }\textbf {\bibinfo {volume} {404}},\
  \bibinfo {pages} {289} (\bibinfo {year} {2000})}\BibitemShut {NoStop}%
\bibitem [{\citenamefont {Tumin}\ and\ \citenamefont
  {Reshotko}(2003)}]{Tumin2}%
  \BibitemOpen
  \bibfield  {author} {\bibinfo {author} {\bibfnamefont {A.}~\bibnamefont
  {Tumin}}\ and\ \bibinfo {author} {\bibfnamefont {E.}~\bibnamefont
  {Reshotko}},\ }\bibfield  {title} {\enquote {\bibinfo {title} {Optimal
  disturbances in compressible boundary layers},}\ }\href@noop {} {\bibfield
  {journal} {\bibinfo  {journal} {AIAA J.}\ }\textbf {\bibinfo {volume} {41}},\
  \bibinfo {pages} {2357} (\bibinfo {year} {2003})}\BibitemShut {NoStop}%
\bibitem [{\citenamefont {Zuccher}\ \emph {et~al.}(2006)\citenamefont
  {Zuccher}, \citenamefont {Tumin},\ and\ \citenamefont {Reshotko}}]{Zuccher}%
  \BibitemOpen
  \bibfield  {author} {\bibinfo {author} {\bibfnamefont {S.}~\bibnamefont
  {Zuccher}}, \bibinfo {author} {\bibfnamefont {A.}~\bibnamefont {Tumin}}, \
  and\ \bibinfo {author} {\bibfnamefont {E.}~\bibnamefont {Reshotko}},\
  }\bibfield  {title} {\enquote {\bibinfo {title} {Parabolic approach to
  optimal perturbation in compressible boundary layers},}\ }\href@noop {}
  {\bibfield  {journal} {\bibinfo  {journal} {J. Fluid Mech.}\ }\textbf
  {\bibinfo {volume} {556}},\ \bibinfo {pages} {189} (\bibinfo {year}
  {2006})}\BibitemShut {NoStop}%
\bibitem [{\citenamefont {Tempelmann}\ \emph {et~al.}(2012)\citenamefont
  {Tempelmann}, \citenamefont {Hanifi},\ and\ \citenamefont
  {Henningson}}]{Tempelmann2}%
  \BibitemOpen
  \bibfield  {author} {\bibinfo {author} {\bibfnamefont {D.}~\bibnamefont
  {Tempelmann}}, \bibinfo {author} {\bibfnamefont {A.}~\bibnamefont {Hanifi}},
  \ and\ \bibinfo {author} {\bibfnamefont {D.~S.}\ \bibnamefont {Henningson}},\
  }\bibfield  {title} {\enquote {\bibinfo {title} {Spatial optimal growth in
  three-dimensional compressible boundary layers},}\ }\href@noop {} {\bibfield
  {journal} {\bibinfo  {journal} {J. Fluid Mech.}\ }\textbf {\bibinfo {volume}
  {704}},\ \bibinfo {pages} {251} (\bibinfo {year} {2012})}\BibitemShut
  {NoStop}%
\bibitem [{\citenamefont {Paredes}\ \emph {et~al.}(2016)\citenamefont
  {Paredes}, \citenamefont {Choudhari}, \citenamefont {Li},\ and\ \citenamefont
  {Chang}}]{Paredes1}%
  \BibitemOpen
  \bibfield  {author} {\bibinfo {author} {\bibfnamefont {P.}~\bibnamefont
  {Paredes}}, \bibinfo {author} {\bibfnamefont {M.~M.}\ \bibnamefont
  {Choudhari}}, \bibinfo {author} {\bibfnamefont {F.}~\bibnamefont {Li}}, \
  and\ \bibinfo {author} {\bibfnamefont {C.-L.}\ \bibnamefont {Chang}},\
  }\bibfield  {title} {\enquote {\bibinfo {title} {Transient growth analysis of
  compressible boundary layers with parabolized stability equations},}\ }in\
  \href@noop {} {\emph {\bibinfo {booktitle} {Proceedings of the 54th AIAA
  Aerospace Sciences Meeting}}}\ (\bibinfo {address} {San Diego, CA},\ \bibinfo
  {year} {2016})\ p.\ \bibinfo {pages} {0051 (19 pages)}\BibitemShut {NoStop}%
\bibitem [{\citenamefont {Dwivedi}\ \emph {et~al.}(2017)\citenamefont
  {Dwivedi}, \citenamefont {Nichols}, \citenamefont {Jovanovi\'c},\ and\
  \citenamefont {Candler}}]{Dwivedi1}%
  \BibitemOpen
  \bibfield  {author} {\bibinfo {author} {\bibfnamefont {A.}~\bibnamefont
  {Dwivedi}}, \bibinfo {author} {\bibfnamefont {J.~W.}\ \bibnamefont
  {Nichols}}, \bibinfo {author} {\bibfnamefont {M.~R.}\ \bibnamefont
  {Jovanovi\'c}}, \ and\ \bibinfo {author} {\bibfnamefont {G.~V.}\ \bibnamefont
  {Candler}},\ }\bibfield  {title} {\enquote {\bibinfo {title} {Optimal spatial
  growth of streaks in oblique shock/boundary layer interaction},}\ }in\
  \href@noop {} {\emph {\bibinfo {booktitle} {Proceedings of the 8th AIAA
  Theoretical Fluid Mechanics Conference}}}\ (\bibinfo {address} {Denver, CO},\
  \bibinfo {year} {2017})\ p.\ \bibinfo {pages} {4163}\BibitemShut {NoStop}%
\bibitem [{\citenamefont {Semper}\ \emph {et~al.}(2012)\citenamefont {Semper},
  \citenamefont {Pruski},\ and\ \citenamefont {Bowersox}}]{Semper}%
  \BibitemOpen
  \bibfield  {author} {\bibinfo {author} {\bibfnamefont {M.~T.}\ \bibnamefont
  {Semper}}, \bibinfo {author} {\bibfnamefont {B.~J.}\ \bibnamefont {Pruski}},
  \ and\ \bibinfo {author} {\bibfnamefont {R.~D.~W.}\ \bibnamefont
  {Bowersox}},\ }\bibfield  {title} {\enquote {\bibinfo {title} {Freestream
  turbulence measurements in a continuously variable hypersonic wind tunnel},}\
  }in\ \href@noop {} {\emph {\bibinfo {booktitle} {Proceedings of the 50th AIAA
  Aerospace Sciences Meeting}}}\ (\bibinfo {address} {Nashville, Tennessee},\
  \bibinfo {year} {2012})\ p.\ \bibinfo {pages} {0732 (13 pages)}\BibitemShut
  {NoStop}%
\bibitem [{\citenamefont {Sesterhenn}(2000)}]{Sesterhenn}%
  \BibitemOpen
  \bibfield  {author} {\bibinfo {author} {\bibfnamefont {J.}~\bibnamefont
  {Sesterhenn}},\ }\bibfield  {title} {\enquote {\bibinfo {title} {A
  characteristic-type formulation of the navier-stokes equations for high order
  upwind schemes},}\ }\href@noop {} {\bibfield  {journal} {\bibinfo  {journal}
  {Comput. Fluids}\ }\textbf {\bibinfo {volume} {30}},\ \bibinfo {pages} {37}
  (\bibinfo {year} {2000})}\BibitemShut {NoStop}%
\bibitem [{\citenamefont {Nichols}\ and\ \citenamefont
  {Lele}(2011{\natexlab{a}})}]{Nichols2}%
  \BibitemOpen
  \bibfield  {author} {\bibinfo {author} {\bibfnamefont {J.~W.}\ \bibnamefont
  {Nichols}}\ and\ \bibinfo {author} {\bibfnamefont {S.~K.}\ \bibnamefont
  {Lele}},\ }\bibfield  {title} {\enquote {\bibinfo {title} {Global modes and
  transient response of a cold supersonic jet},}\ }\href@noop {} {\bibfield
  {journal} {\bibinfo  {journal} {J. Fluid Mech.}\ }\textbf {\bibinfo {volume}
  {669}},\ \bibinfo {pages} {225} (\bibinfo {year}
  {2011}{\natexlab{a}})}\BibitemShut {NoStop}%
\bibitem [{\citenamefont {Hildebrand}\ \emph
  {et~al.}(2018{\natexlab{b}})\citenamefont {Hildebrand}, \citenamefont
  {Nichols}, \citenamefont {Candler},\ and\ \citenamefont
  {Jovanovi\'c}}]{Hildebrand_CITE}%
  \BibitemOpen
  \bibfield  {author} {\bibinfo {author} {\bibfnamefont {N.}~\bibnamefont
  {Hildebrand}}, \bibinfo {author} {\bibfnamefont {J.~W.}\ \bibnamefont
  {Nichols}}, \bibinfo {author} {\bibfnamefont {G.~V.}\ \bibnamefont
  {Candler}}, \ and\ \bibinfo {author} {\bibfnamefont {M.~R.}\ \bibnamefont
  {Jovanovi\'c}},\ }\bibfield  {title} {\enquote {\bibinfo {title} {Transient
  growth in oblique shock wave/laminar boundary layer interactions at {M}ach
  5.92},}\ }in\ \href@noop {} {\emph {\bibinfo {booktitle} {Proceedings of the
  48th AIAA Fluid Dynamics Conference}}}\ (\bibinfo {address} {Atlanta, GA},\
  \bibinfo {year} {2018})\ p.\ \bibinfo {pages} {3221 (17 pages)}\BibitemShut
  {NoStop}%
\bibitem [{\citenamefont {Luchini}\ and\ \citenamefont
  {Bottaro}(2013)}]{Luchini3}%
  \BibitemOpen
  \bibfield  {author} {\bibinfo {author} {\bibfnamefont {P.}~\bibnamefont
  {Luchini}}\ and\ \bibinfo {author} {\bibfnamefont {A.}~\bibnamefont
  {Bottaro}},\ }\bibfield  {title} {\enquote {\bibinfo {title} {Adjoint
  equations in stability analysis},}\ }\href@noop {} {\bibfield  {journal}
  {\bibinfo  {journal} {Annu. Rev. Fluid Mech.}\ }\textbf {\bibinfo {volume}
  {46}},\ \bibinfo {pages} {493} (\bibinfo {year} {2013})}\BibitemShut
  {NoStop}%
\bibitem [{\citenamefont {Chu}(1965)}]{Chu}%
  \BibitemOpen
  \bibfield  {author} {\bibinfo {author} {\bibfnamefont {B.-T.}\ \bibnamefont
  {Chu}},\ }\bibfield  {title} {\enquote {\bibinfo {title} {On the energy
  transfer to small disturbances in fluid flow (part {I})},}\ }\href@noop {}
  {\bibfield  {journal} {\bibinfo  {journal} {Acta Mech.}\ }\textbf {\bibinfo
  {volume} {1}},\ \bibinfo {pages} {215} (\bibinfo {year} {1965})}\BibitemShut
  {NoStop}%
\bibitem [{\citenamefont {Gu\'egan}\ \emph {et~al.}(2006)\citenamefont
  {Gu\'egan}, \citenamefont {Schmid},\ and\ \citenamefont {Huerre}}]{Guegan}%
  \BibitemOpen
  \bibfield  {author} {\bibinfo {author} {\bibfnamefont {A.}~\bibnamefont
  {Gu\'egan}}, \bibinfo {author} {\bibfnamefont {P.~J.}\ \bibnamefont
  {Schmid}}, \ and\ \bibinfo {author} {\bibfnamefont {P.}~\bibnamefont
  {Huerre}},\ }\bibfield  {title} {\enquote {\bibinfo {title} {Optimal energy
  growth and optimal control in swept hiemenz flow},}\ }\href@noop {}
  {\bibfield  {journal} {\bibinfo  {journal} {J. Fluid Mech.}\ }\textbf
  {\bibinfo {volume} {566}},\ \bibinfo {pages} {11} (\bibinfo {year}
  {2006})}\BibitemShut {NoStop}%
\bibitem [{\citenamefont {Candler}\ \emph
  {et~al.}(2015{\natexlab{a}})\citenamefont {Candler}, \citenamefont {Johnson},
  \citenamefont {Nompelis}, \citenamefont {Gidzak}, \citenamefont
  {Subbareddy},\ and\ \citenamefont {Barnhardt}}]{candler2015development}%
  \BibitemOpen
  \bibfield  {author} {\bibinfo {author} {\bibfnamefont {G.~V.}\ \bibnamefont
  {Candler}}, \bibinfo {author} {\bibfnamefont {H.~B.}\ \bibnamefont
  {Johnson}}, \bibinfo {author} {\bibfnamefont {I.}~\bibnamefont {Nompelis}},
  \bibinfo {author} {\bibfnamefont {V.~M.}\ \bibnamefont {Gidzak}}, \bibinfo
  {author} {\bibfnamefont {P.~K.}\ \bibnamefont {Subbareddy}}, \ and\ \bibinfo
  {author} {\bibfnamefont {M.}~\bibnamefont {Barnhardt}},\ }\bibfield  {title}
  {\enquote {\bibinfo {title} {Development of the us3d code for advanced
  compressible and reacting flow simulations},}\ }in\ \href@noop {} {\emph
  {\bibinfo {booktitle} {53rd AIAA Aerospace Sciences Meeting}}}\ (\bibinfo
  {address} {Kissimmee, FL},\ \bibinfo {year} {2015})\ p.\ \bibinfo {pages}
  {1893 (26 pages)}\BibitemShut {NoStop}%
\bibitem [{\citenamefont {Li}\ and\ \citenamefont {Demmel}(2003)}]{Li}%
  \BibitemOpen
  \bibfield  {author} {\bibinfo {author} {\bibfnamefont {X.~S.}\ \bibnamefont
  {Li}}\ and\ \bibinfo {author} {\bibfnamefont {J.~W.}\ \bibnamefont
  {Demmel}},\ }\bibfield  {title} {\enquote {\bibinfo {title}
  {Super{LU}$\_$dist: A scalable distributed memory sparse direct solver for
  unsymmetric linear systems},}\ }\href@noop {} {\bibfield  {journal} {\bibinfo
   {journal} {ACM Trans. Math. Softw. ({TOMS})}\ }\textbf {\bibinfo {volume}
  {29}},\ \bibinfo {pages} {110} (\bibinfo {year} {2003})}\BibitemShut
  {NoStop}%
\bibitem [{\citenamefont {Durran}(1999)}]{Durran}%
  \BibitemOpen
  \bibfield  {author} {\bibinfo {author} {\bibfnamefont {D.~R.}\ \bibnamefont
  {Durran}},\ }\href@noop {} {\emph {\bibinfo {title} {Numerical methods for
  wave equations in geophysical fluid dynamics}}}\ (\bibinfo  {publisher}
  {Springer-Verlag, New York},\ \bibinfo {year} {1999})\BibitemShut {NoStop}%
\bibitem [{\citenamefont {Mani}(2010)}]{Mani}%
  \BibitemOpen
  \bibfield  {author} {\bibinfo {author} {\bibfnamefont {A.}~\bibnamefont
  {Mani}},\ }\bibfield  {title} {\enquote {\bibinfo {title} {On the
  reflectivity of sponge zones in compressible flow simulations},}\ }in\
  \href@noop {} {\emph {\bibinfo {booktitle} {Annual Research Briefs, Center
  for Turbulence Research}}}\ (\bibinfo {address} {Stanford University,
  Stanford, CA},\ \bibinfo {year} {2010})\ pp.\ \bibinfo {pages}
  {117--133}\BibitemShut {NoStop}%
\bibitem [{\citenamefont {Landhal}(1977)}]{Landahl1}%
  \BibitemOpen
  \bibfield  {author} {\bibinfo {author} {\bibfnamefont {M.~T.}\ \bibnamefont
  {Landhal}},\ }\bibfield  {title} {\enquote {\bibinfo {title} {Dynamics of
  boundary layer turbulence and the mechanism of drag reduction},}\ }\href@noop
  {} {\bibfield  {journal} {\bibinfo  {journal} {Phys. Fluids}\ }\textbf
  {\bibinfo {volume} {20}},\ \bibinfo {pages} {S55} (\bibinfo {year}
  {1977})}\BibitemShut {NoStop}%
\bibitem [{\citenamefont {Landhal}(1980)}]{Landahl2}%
  \BibitemOpen
  \bibfield  {author} {\bibinfo {author} {\bibfnamefont {M.~T.}\ \bibnamefont
  {Landhal}},\ }\bibfield  {title} {\enquote {\bibinfo {title} {A note on an
  algebraic instability of inviscid parallel shear flows},}\ }\href@noop {}
  {\bibfield  {journal} {\bibinfo  {journal} {J. Fluid Mech.}\ }\textbf
  {\bibinfo {volume} {98}},\ \bibinfo {pages} {243} (\bibinfo {year}
  {1980})}\BibitemShut {NoStop}%
\bibitem [{\citenamefont {Shrestha}\ \emph {et~al.}(2016)\citenamefont
  {Shrestha}, \citenamefont {Dwivedi}, \citenamefont {Hildebrand},
  \citenamefont {Nichols}, \citenamefont {Jovanovi\'c},\ and\ \citenamefont
  {Candler}}]{Shrestha}%
  \BibitemOpen
  \bibfield  {author} {\bibinfo {author} {\bibfnamefont {P.}~\bibnamefont
  {Shrestha}}, \bibinfo {author} {\bibfnamefont {A.}~\bibnamefont {Dwivedi}},
  \bibinfo {author} {\bibfnamefont {N.}~\bibnamefont {Hildebrand}}, \bibinfo
  {author} {\bibfnamefont {J.~W.}\ \bibnamefont {Nichols}}, \bibinfo {author}
  {\bibfnamefont {M.~R.}\ \bibnamefont {Jovanovi\'c}}, \ and\ \bibinfo {author}
  {\bibfnamefont {G.~V.}\ \bibnamefont {Candler}},\ }\bibfield  {title}
  {\enquote {\bibinfo {title} {Interaction of an oblique shock with a
  transitional {M}ach $5.65$ boundary layer},}\ }in\ \href@noop {} {\emph
  {\bibinfo {booktitle} {Proceedings of the 46th AIAA Fluid Dynamics
  Conference}}}\ (\bibinfo {address} {Washington, DC},\ \bibinfo {year}
  {2016})\ p.\ \bibinfo {pages} {3647 (14 pages)}\BibitemShut {NoStop}%
\bibitem [{\citenamefont {Cossu}\ and\ \citenamefont {Chomaz}(1997)}]{Cossu}%
  \BibitemOpen
  \bibfield  {author} {\bibinfo {author} {\bibfnamefont {C.}~\bibnamefont
  {Cossu}}\ and\ \bibinfo {author} {\bibfnamefont {J.-M.}\ \bibnamefont
  {Chomaz}},\ }\bibfield  {title} {\enquote {\bibinfo {title} {Global measures
  of local convective instabilities},}\ }\href@noop {} {\bibfield  {journal}
  {\bibinfo  {journal} {Phys. Rev. Lett.}\ }\textbf {\bibinfo {volume} {78}},\
  \bibinfo {pages} {4387 (4 pages)} (\bibinfo {year} {1997})}\BibitemShut
  {NoStop}%
\bibitem [{\citenamefont {Nichols}\ and\ \citenamefont
  {Lele}(2011{\natexlab{b}})}]{Nichols4}%
  \BibitemOpen
  \bibfield  {author} {\bibinfo {author} {\bibfnamefont {J.~W.}\ \bibnamefont
  {Nichols}}\ and\ \bibinfo {author} {\bibfnamefont {S.~K.}\ \bibnamefont
  {Lele}},\ }\bibfield  {title} {\enquote {\bibinfo {title} {Non-normal global
  modes of high-speed jets},}\ }\href@noop {} {\bibfield  {journal} {\bibinfo
  {journal} {Int. J. Spray Combust. Dyn.}\ }\textbf {\bibinfo {volume} {3}},\
  \bibinfo {pages} {285} (\bibinfo {year} {2011}{\natexlab{b}})}\BibitemShut
  {NoStop}%
\bibitem [{\citenamefont {Tempelmann}\ \emph {et~al.}(2010)\citenamefont
  {Tempelmann}, \citenamefont {Hanifi},\ and\ \citenamefont
  {Henningson}}]{Tempelmann1}%
  \BibitemOpen
  \bibfield  {author} {\bibinfo {author} {\bibfnamefont {D.}~\bibnamefont
  {Tempelmann}}, \bibinfo {author} {\bibfnamefont {A.}~\bibnamefont {Hanifi}},
  \ and\ \bibinfo {author} {\bibfnamefont {D.~S.}\ \bibnamefont {Henningson}},\
  }\bibfield  {title} {\enquote {\bibinfo {title} {Spatial optimal growth in
  three-dimensional boundary layers},}\ }\href@noop {} {\bibfield  {journal}
  {\bibinfo  {journal} {J. Fluid Mech.}\ }\textbf {\bibinfo {volume} {646}},\
  \bibinfo {pages} {5} (\bibinfo {year} {2010})}\BibitemShut {NoStop}%
\bibitem [{\citenamefont {Candler}\ \emph
  {et~al.}(2015{\natexlab{b}})\citenamefont {Candler}, \citenamefont
  {Subbareddy},\ and\ \citenamefont {Nompelis}}]{Candler}%
  \BibitemOpen
  \bibfield  {author} {\bibinfo {author} {\bibfnamefont {G.~V.}\ \bibnamefont
  {Candler}}, \bibinfo {author} {\bibfnamefont {P.~K.}\ \bibnamefont
  {Subbareddy}}, \ and\ \bibinfo {author} {\bibfnamefont {I.}~\bibnamefont
  {Nompelis}},\ }\bibfield  {title} {\enquote {\bibinfo {title} {C{FD} methods
  for hypersonic flows and aerothermodynamics},}\ }in\ \href@noop {} {\emph
  {\bibinfo {booktitle} {Hypersonic Nonequilibrium Flows: Fundamentals and
  Recent Advances}}},\ \bibinfo {editor} {edited by\ \bibinfo {editor}
  {\bibfnamefont {E.}~\bibnamefont {Josyula}}}\ (\bibinfo  {publisher} {AIAA},\
  \bibinfo {year} {2015})\ pp.\ \bibinfo {pages} {203--237}\BibitemShut
  {NoStop}%
\bibitem [{\citenamefont {Alizard}\ \emph {et~al.}(2008)\citenamefont
  {Alizard}, \citenamefont {Cherubini},\ and\ \citenamefont
  {Robinet}}]{Alizard}%
  \BibitemOpen
  \bibfield  {author} {\bibinfo {author} {\bibfnamefont {F.}~\bibnamefont
  {Alizard}}, \bibinfo {author} {\bibfnamefont {S.}~\bibnamefont {Cherubini}},
  \ and\ \bibinfo {author} {\bibfnamefont {J.-C.}\ \bibnamefont {Robinet}},\
  }\bibfield  {title} {\enquote {\bibinfo {title} {Sensitivity and optimal
  forcing response in separated boundary layer flows},}\ }\href@noop {}
  {\bibfield  {journal} {\bibinfo  {journal} {Phys. Fluids}\ }\textbf {\bibinfo
  {volume} {21}},\ \bibinfo {pages} {064108 (13 pages)} (\bibinfo {year}
  {2008})}\BibitemShut {NoStop}%
\bibitem [{\citenamefont {Cherubini}\ \emph {et~al.}(2010)\citenamefont
  {Cherubini}, \citenamefont {Robinet},\ and\ \citenamefont
  {Palma}}]{Cherubini}%
  \BibitemOpen
  \bibfield  {author} {\bibinfo {author} {\bibfnamefont {S.}~\bibnamefont
  {Cherubini}}, \bibinfo {author} {\bibfnamefont {J.-C.}\ \bibnamefont
  {Robinet}}, \ and\ \bibinfo {author} {\bibfnamefont {P.~D.}\ \bibnamefont
  {Palma}},\ }\bibfield  {title} {\enquote {\bibinfo {title} {The effects of
  non-normality and nonlinearity of the navier–stokes operator on the
  dynamics of a large laminar separation bubble},}\ }\href@noop {} {\bibfield
  {journal} {\bibinfo  {journal} {Phys. Fluids}\ }\textbf {\bibinfo {volume}
  {22}},\ \bibinfo {pages} {014102 (15 pages)} (\bibinfo {year}
  {2010})}\BibitemShut {NoStop}%
\bibitem [{\citenamefont {$\mathring{\mathrm{A}}$kervik}\ \emph
  {et~al.}(2007)\citenamefont {$\mathring{\mathrm{A}}$kervik}, \citenamefont
  {H\oe{}pffner}, \citenamefont {Ehrenstein},\ and\ \citenamefont
  {Henningson}}]{Akervik2}%
  \BibitemOpen
  \bibfield  {author} {\bibinfo {author} {\bibfnamefont {E.}~\bibnamefont
  {$\mathring{\mathrm{A}}$kervik}}, \bibinfo {author} {\bibfnamefont
  {J.}~\bibnamefont {H\oe{}pffner}}, \bibinfo {author} {\bibfnamefont
  {U.}~\bibnamefont {Ehrenstein}}, \ and\ \bibinfo {author} {\bibfnamefont
  {D.~S.}\ \bibnamefont {Henningson}},\ }\bibfield  {title} {\enquote {\bibinfo
  {title} {Optimal growth, model reduction and control in a separated
  boundary-layer flow using global eigenmodes},}\ }\href@noop {} {\bibfield
  {journal} {\bibinfo  {journal} {J. Fluid Mech.}\ }\textbf {\bibinfo {volume}
  {579}},\ \bibinfo {pages} {305} (\bibinfo {year} {2007})}\BibitemShut
  {NoStop}%
\bibitem [{\citenamefont {$\mathring{\mathrm{A}}$kervik}\ \emph
  {et~al.}(2008)\citenamefont {$\mathring{\mathrm{A}}$kervik}, \citenamefont
  {Ehrenstein}, \citenamefont {Gallaire},\ and\ \citenamefont
  {Henningson}}]{Akervik1}%
  \BibitemOpen
  \bibfield  {author} {\bibinfo {author} {\bibfnamefont {E.}~\bibnamefont
  {$\mathring{\mathrm{A}}$kervik}}, \bibinfo {author} {\bibfnamefont
  {U.}~\bibnamefont {Ehrenstein}}, \bibinfo {author} {\bibfnamefont
  {F.}~\bibnamefont {Gallaire}}, \ and\ \bibinfo {author} {\bibfnamefont
  {D.~S.}\ \bibnamefont {Henningson}},\ }\bibfield  {title} {\enquote {\bibinfo
  {title} {Global two-dimensional stability measures of the flat plate boundary
  layer flow},}\ }\href@noop {} {\bibfield  {journal} {\bibinfo  {journal}
  {Eur. J. Mech.}\ }\textbf {\bibinfo {volume} {27}},\ \bibinfo {pages} {501}
  (\bibinfo {year} {2008})}\BibitemShut {NoStop}%
\bibitem [{\citenamefont {Chomaz}(2005)}]{Chomaz}%
  \BibitemOpen
  \bibfield  {author} {\bibinfo {author} {\bibfnamefont {J.-M.}\ \bibnamefont
  {Chomaz}},\ }\bibfield  {title} {\enquote {\bibinfo {title} {Global
  instabilities in spatially developing flows: non-normality and
  nonlinearity},}\ }\href@noop {} {\bibfield  {journal} {\bibinfo  {journal}
  {Annu. Rev. Fluid Mech.}\ }\textbf {\bibinfo {volume} {37}},\ \bibinfo
  {pages} {357} (\bibinfo {year} {2005})}\BibitemShut {NoStop}%
\bibitem [{\citenamefont {Gallaire}\ and\ \citenamefont
  {Chomaz}(2003)}]{Gallaire}%
  \BibitemOpen
  \bibfield  {author} {\bibinfo {author} {\bibfnamefont {F.}~\bibnamefont
  {Gallaire}}\ and\ \bibinfo {author} {\bibfnamefont {J.-M.}\ \bibnamefont
  {Chomaz}},\ }\bibfield  {title} {\enquote {\bibinfo {title} {Mode selection
  in swirling jet experiments: a linear stability analysis},}\ }\href@noop {}
  {\bibfield  {journal} {\bibinfo  {journal} {J. Fluid Mech.}\ }\textbf
  {\bibinfo {volume} {494}},\ \bibinfo {pages} {223} (\bibinfo {year}
  {2003})}\BibitemShut {NoStop}%
\bibitem [{\citenamefont {Delbende}\ \emph {et~al.}(1998)\citenamefont
  {Delbende}, \citenamefont {Chomaz},\ and\ \citenamefont {Huerre}}]{Delbende}%
  \BibitemOpen
  \bibfield  {author} {\bibinfo {author} {\bibfnamefont {I.}~\bibnamefont
  {Delbende}}, \bibinfo {author} {\bibfnamefont {J.-M.}\ \bibnamefont
  {Chomaz}}, \ and\ \bibinfo {author} {\bibfnamefont {P.}~\bibnamefont
  {Huerre}},\ }\bibfield  {title} {\enquote {\bibinfo {title}
  {Absolute/convective instabilities in the batchelor vortex: a numerical study
  of the linear impulse response},}\ }\href@noop {} {\bibfield  {journal}
  {\bibinfo  {journal} {J. Fluid Mech.}\ }\textbf {\bibinfo {volume} {355}},\
  \bibinfo {pages} {229} (\bibinfo {year} {1998})}\BibitemShut {NoStop}%
\end{thebibliography}%

%merlin.mbs apsrev4-1.bst 2010-07-25 4.21a (PWD, AO, DPC) hacked
%Control: key (0)
%Control: author (72) initials jnrlst
%Control: editor formatted (1) identically to author
%Control: production of article title (1) required
%Control: page (0) single
%Control: year (1) truncated
%Control: production of eprint (0) enabled
%

\end{document}